\documentclass[useAMS,usenatbib]{mn2e}
\usepackage{graphicx}

\title[Oscillations of Thick Discs Around BHs -- II] {Oscillations of
  Thick Accretion Discs \\ Around Black Holes -- II}
  \author[E. Rubio-Herrera \& W. H. Lee] {Eduardo Rubio-Herrera
  \thanks{E--mail: eduardo@astroscu.unam.mx} and William H.~ Lee \\
  Instituto de Astronom\'{\i}a, UNAM, Apartado Postal 70--264
  C.P. 04510 M\'exico D.F. M\'exico. \\} \date{Released 2005 Xxxxx XX}

\pagerange{\pageref{firstpage}--\pageref{lastpage}} \pubyear{2005}

\def\LaTeX{L\kern-.36em\raise.3ex\hbox{a}\kern-.15em
    T\kern-.1667em\lower.7ex\hbox{E}\kern-.125emX}

\begin{document}

\label{firstpage}

\maketitle

\begin{abstract}
We present a numerical study of the global modes of oscillation of
thick accretion discs around black holes. We have previously studied
the case of constant distributions of specific angular momentum. In
this second paper, we investigate (i) how the size of the disc affects
the oscillation eigenfrequencies, and (ii) the effect of power--law
distributions of angular momentum on the oscillations. In particular,
we compare the oscillations of the disc with the epicyclic
eigenfrequencies of a test particle with different angular momentum
distributions orbiting around the central object.  We find that there
is a frequency shift away from the epicyclic eigenfrequency of the
test particle to lower values as the size of the tori is increased.
We have also studied the response of a thick accretion disc to a
localized external perturbation using non constant specific angular
momentum distributions within the disc. We find that in this case it
is also possible (as reported previously for constant angular momentum
distributions) to efficiently excite internal modes of oscillation. In
fact we show here that the local perturbations excite global
oscillations (acoustic p modes) closely related to the epicyclic
oscillations of test particles. Our results are particularly relevant
in the context of low mass X--ray binaries and microquasars, and the
high frequency Quasi--Periodic Oscillations (QPOs) observed in them.
Our computations make use of a Smooth Particle Hydrodynamics (SPH)
code in azimuthal symmetry, and use a gravitational potential that
mimics the effects of strong gravity.
\end{abstract}


\section{Introduction}\label{intro}
Thick accretion discs may appear in systems containing compact objects
such as black holes and neutron stars.  These could form in different
scenarios governed by gravitational forces.  Specific examples are:
(i) gravitational collapse of a rotating massive star, (ii) the
coalescence of two compact objects or (iii) accretion phenomena
asociated with Low Mass X--ray Binaries (LMXBs).

The first case (gravitational collapse), may be related to the
production of $\gamma$-ray bursts (GRBs) as in the collapsar or
hypernova models. The collapsar (or failed supernova) was proposed by
\citet{w93} and offers the possibility to form a thick disc when the
mantle of a rotating massive star ($M \sim 25 M_{\odot}$) falls freely
onto the recently formed black hole. The material has a very high
angular momentum, which prevents it from directly accreting and a
centrifugally supported disc is formed. These systems are related to
energetic, long duration ($> 2$ s) GRBs accompanied by a Type Ib$/$Ic
supernova \citep{ketal98,setal03,hjorth03}.  Similar numerical
relativistic calculations concerning the collapse of massive stars,
reported by \citet{sasi02} and \citet{sisa02} may explain the
formation of massive black holes surounded by a debris torus. The
hypernova scenario was proposed by \citet{p98} and is also related to
GRBs.  Hypernov\ae~ are extremely energetic supernova explosions (\emph{kinetic  energy}
$\sim 3 \times 10^{52}$ erg), which consider (a) a single star
(i.e. Wolf--Rayet type star) or (b) a binary system formed by massive
stars ($M \sim 20-25 M_{\odot}$) orbiting closely ($\sim
1.5R_{\odot}$) around each other, and which eventually merge to form a
single star [see \citet{no04} for a comprehensive review].  This star
will undergo gravitational collapse, leading to the formation of a
compact object. Due to the rapid rotation, a thick accretion disc
around the remaining black hole will be formed. In either case, a hot,
thick disc of $\sim 0.1 M_{\odot}$ is expected around the newborn
black hole.

Another posibility to form thick discs is via the coalescence of two
compact objects, leading to the formation of a system composed of a
black hole and an accretion torus.  This has been investigated by
various groups using nummerical simulations. Some of the first results
were obtained by \citet{da94}, who modeled the merger of two neutron
stars. They showed that after few orbital periods the residual
material could form a thick disc. These results have been confirmed
and extended \citep[see e.g.,][]{rjs96,rj99} for the merger of two
neutron stars, where again, a hot thick disc is formed around a
central mass, which will presumably collapse to a black
hole. Investigating the more complex fusion of a neutron star and a
black hole, \citet{kl98} and \citet{lee} found that after the initial
mass transfer there is a remnant of material around the black hole
\citep[see also recent results by][]{rsw04}.  All these scenarios may
form massive ($M_d/M_{BH} \approx 0.1$) and dense discs.

There are also low--mass discs ($M_d/M_{BH} \ll$ 1) which are formed
through mass transfer in a binary system. A steady supply of mass,
energy and angular momentum produces structures which are nearly in a
steady state, but which nevertheless exhibit variability on various
timescales \citep[see e.g.,][]{kato98}. \\

Once the disc is formed, one may ask:~ under which conditions is it
dynamically stable or unstable (quite apart from the question of
thermal stability, if dynamically stable)? Answering this in any
detail requires the consideration of viscosity, self--gravity,
magnetic fields and the angular momentum distribution of the material
within the disc.  This last point led \citet{pp84} to investigate the
stability of an isentropic and constant angular momentum disc, and
find that it is unstable to global non--axisymetric perturbations.
Three--dimensional calculations by ~\citet{zb86} found that there is
indeed a redistribution of angular momentum within the disc, due to
the growth of non-axisymetric instabilities. In the presence of
accretion, however, this instability may be suppressed
\citep{blaes87}, and thick tori with relatively flat distributions of
angular momentum may actually occur in astrophysical systems.

Perhaps one of the most interesting global instabilities is the
so--called runaway radial instability, which was identified by
\citet{acn}. It is due to a violent mass exchange asociated with the
rapidly changing equipotential Roche surfaces of the disc + black hole
system.  It is catastrophic and leads to the complete destruction of
the disc on a dynamical time scale.  It is well known that in binary
systems there is a point called the interior Lagrangian point $L_{1}$,
located at a radius $r_{L_1}$ where there is a balance between
gravitational and centrifugal forces.  For accreting discs, when mass
transfer from the disc to the central object begins the corresponding
Lagrange point, lying between the compact object and the disc, moves
outward. The accretion rate increases rapidly until the disc is
accreted completely.  \citet{wi84} found that the rotation of the
black hole inhibits the runaway radial instability and later
\citet{akl} confirmed this result and reported that the self-gravity
of the disc has a destabilizing effect.  It appears that the most
critical parameter in the runaway radial instability is the specific
angular momentum distribution within the disc.  In the last twenty
five years there has been much work on this issue from
\emph{stationary} (i.e. Schwarzschild metric) to \emph{dynamical}
(i.e. Kerr metric) space--time models.  A complete summary of the work
done appears in the extensive work of \citet{fd1} and \citet{df}.
These authors conclude, as \citet{dm} did earlier, that the
instability is suppresed when the angular momentum follows a power law
distribution with a positive radial gradient.  In this respect studies
made by \citet{rj99} and \citet{lee} concerning compact binary mergers
show that the distribution of angular momentum in post--merger discs
is far from being constant, and thus they are stable in this
context. \\

Our numerical simulations were performed in order to address two
questions: (i) How do thick discs oscillate? and (ii) How does a
localized perturbation affect the global oscillatory behaviour of the
disc?  These are formulated in the particular context in the
millisecond oscillations discovered in X--ray binary sources by {\it
RXTE} \citep{vk}. These systems have compact objects surrounded by an
accretion disc, whose variability could be imprinted on the X--ray
lightcurve and corresponding power spectra.

Disc oscillations have been investigated numerically and analytically
by several groups before, and may be applied to the quasi periodical
oscillations (QPOs) observed in many X--ray sources
\citep{petal97,wagoner99,swo01,osw02}. One specific model \citep{rymz}
accounts for the high frequency QPOs as a result of p--mode
oscillations of an accretion torus orbiting close to the black
hole. \citet{ryz} investigated theoretically how a relativistic torus
oscillates in a stationary Schwarzschild space--time, showing that the
eigenfrequencies of the axisymmetric oscillations correspond to
acoustic p--modes.  Their results shows that the oscillations behaves
like a sound wave globally traped within the disc, with the
eigenfrequencies appearing in the sequence 2:3:4... independently of
the distribution of angular momentum considered.  The same conclusions
are obtained by \citet{zfrm} using numerical simulations for a
dynamical Kerr space--time, concluding that p--mode oscillations in
relativistic tori could explain the high frequency QPOs observed in
the X--ray binaries.  The oscillatory analysis made by Rezzolla et
al. (2003a), Rezzolla et al. (2003b), Montero et al. (2004) and
Zanotti et al. (2005) has shown theoretically that there is a
frequency shifting to lower frequencies as the size of the torus is
increased.  Their analytical work is restricted to height--integrated
structures and globally applied perturbations. We have confirmed this
result more generally here with a different approach.

From the observations in the X--ray band mentioned above, \citet{ak01}
noted in GRO J1655--40, that the two stable peaks that appear in the
power spectrum in the hHz range are in a 3:2 ratio. The same
commeasurability has now been observed in three more sources:
H1743-322, XTE J1550--564 and GRS 1915+105 \citep[see][for a recent
review]{mr04}. This behaviour may also be present in a system
containing a neutron star (Sco X--1) \citep{abbk}, and could be
explained by parametric resonance in a thin disc
\citep{akklr03,re}. These models predict that the frequencies will
show the epicyclic motions of a perturbed flow line in the accretion
disc or a combination of these and a fixed perturbation associated to
the spin of the central object (in the neutron star sources). This
last issue was analysed by \citet{k04}, \citet{lak} and
\citet[][hereafter Paper~I]{rl} showing that it is indeed posible to
excite internal oscillatory modes efficiently by an external driving
agent.

In this paper we show that there is a shifting to lower frequencies as
the thickness of the disc is increased, independently of the
distribution of angular momentum within the torus. We also show that
it is possible to excite internal p--modes in a thick accretion disc,
by means of a localized perturbation which affects only a small
portion of the disc periodically, exciting the strongest modes
apparent in a 3:2 frequency ratio.

We empasize that the results presented here are purely dynamical in a
pseudo--Newtonian gravitational potential which mimics the properties
of the Schwarzschild metric for a stationary space--time. No attempt
has been made to consider magnetic fields, radiation or the effects of
viscosity.  For definiteness, a numerical value of $2.5M_{\odot}$ has
been used throughout the paper. The reader should keep in mind,
though, that distances are always measured in units of the
gravitational radius. As such, our results are scale free, and can be
applied to any black hole mass, with the frequencies scaling as $\nu
\propto 1/M$.

\section{Initial conditions and numerical method}\label{ICmethod}

\subsection{Hydrostatic equilibrium for a thick torus}\label{hydroeq}

We will define a thick torus as that which has similar radial and
vertical extensions, i.e., $L \sim H$, and where $L$ is comparable to
its radial position, $R$. We will neglect self--gravity and
additionally assume azimuthal symmetry.  We have constructed thick
tori using the pseudo--Newtonian gravitational potential proposed by
\citet{pw}, given by:
\begin{equation}
\Phi = \frac{-G M_{BH}}{R-r_g},
\end{equation}
which qualitatively reproduces the behaviour of a test particle in the
Schwarzschild metric, and accounts correctly for the positions of the
marginally stable and marginally bound orbits. In this equation,
$R=\sqrt{r^2+z^2}$ is the radial distance mesured from the central
mass and $r_g =2GM_{BH}/c^2$ denotes the Schwarzschild radius of the
central object.  Using a polytropic equation of state
$P=K\rho^{\gamma}$, the equations of hydrodynamics in cylindrical
coordinates, and neglecting self gravity, we have:
\begin{equation}
\frac{1}{\rho} \nabla P = - \nabla \Phi_{\rm eff}
\end{equation}
where $\Phi_{\rm eff}$ is the effective potential.  Using the
Paczy\'nski--Wiita expression given in (1), substituting it in (2) and
integrating over $r$ we have,
\begin{equation}
\frac{\gamma}{\gamma -1 } \frac{P}{\rho} + \Phi_{\rm eff} + \Phi_0 =
{\rm cst},
\end{equation}
which describes a torus in hydrostatic equilibrium.  The term $\Phi_0$
can be interpreted as a filling factor of the effective potential
well, and will be discussed below.  Using an arbitrary distribution of
angular momentum $\ell(r)$, which depends only of the position
coordinate $r$ according to the von Zeipel theorem for a polytropic
equation of state, the effective potential takes the form:
\begin{equation}
\Phi_{\rm eff} = \frac{-G M_{BH}}{R-r_g} + \int
\frac{\ell(r')^2}{2r'^3}dr.
\end{equation}
The boundary of the tours is the surface over which the pressure is
zero, and is given by:
\begin{equation}
z= \left\{\left[GM_{BH}\left(\int \frac{ \ell(r')^2}{2r'^3} +\Phi_0
\right)^{-1}+r_g \right]^2 -r^2\right\}^{1/2}.
\end{equation}
The corresponding meridional cross sections over one half the $r$--$z$
plane are shown in Figure~1 for various filling factors $\Phi_{0}$ and
different distributions of specific angular momentum.
\begin{figure}
\includegraphics[width=84mm,height=60mm]{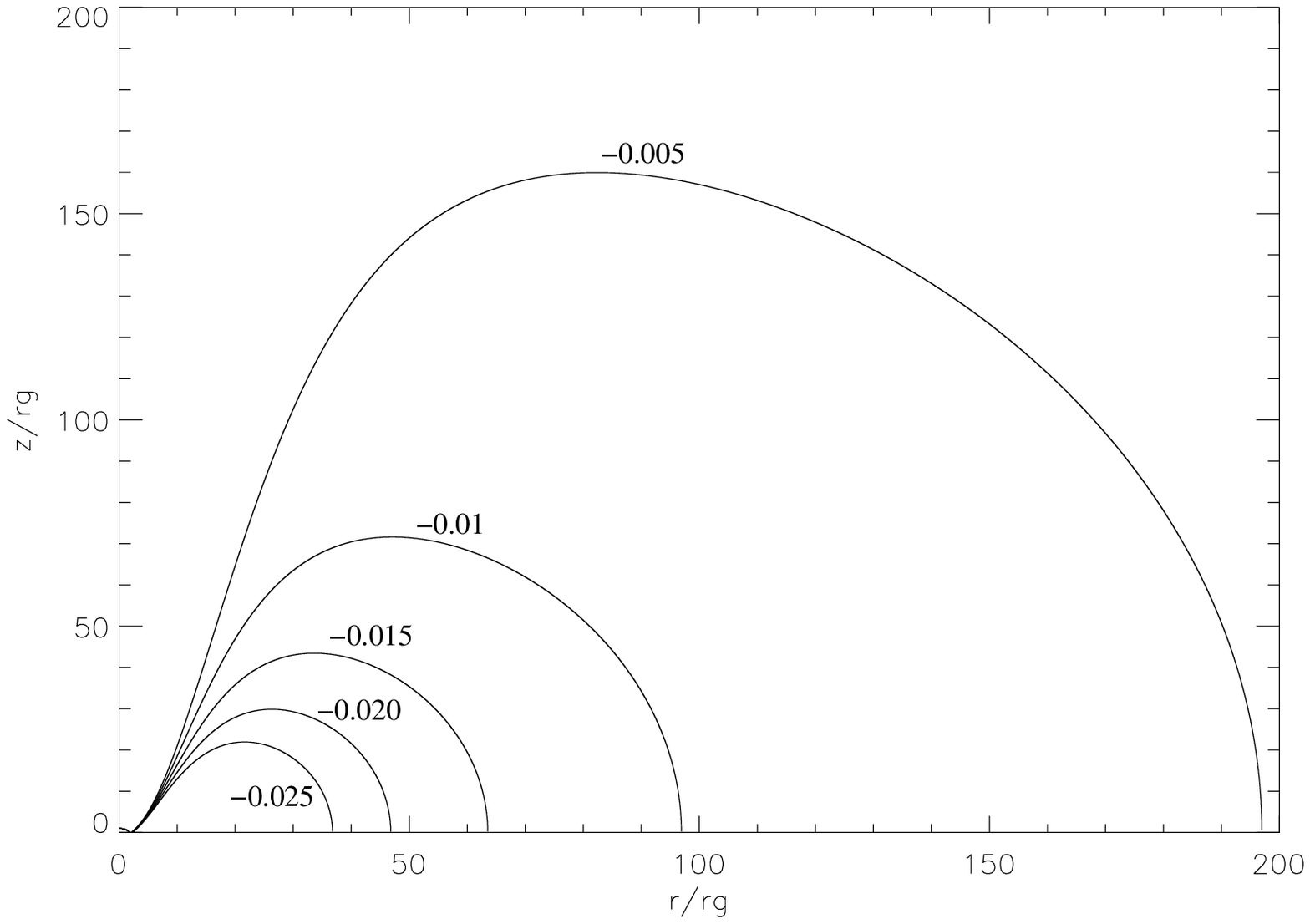}
\includegraphics[width=84mm,height=60mm]{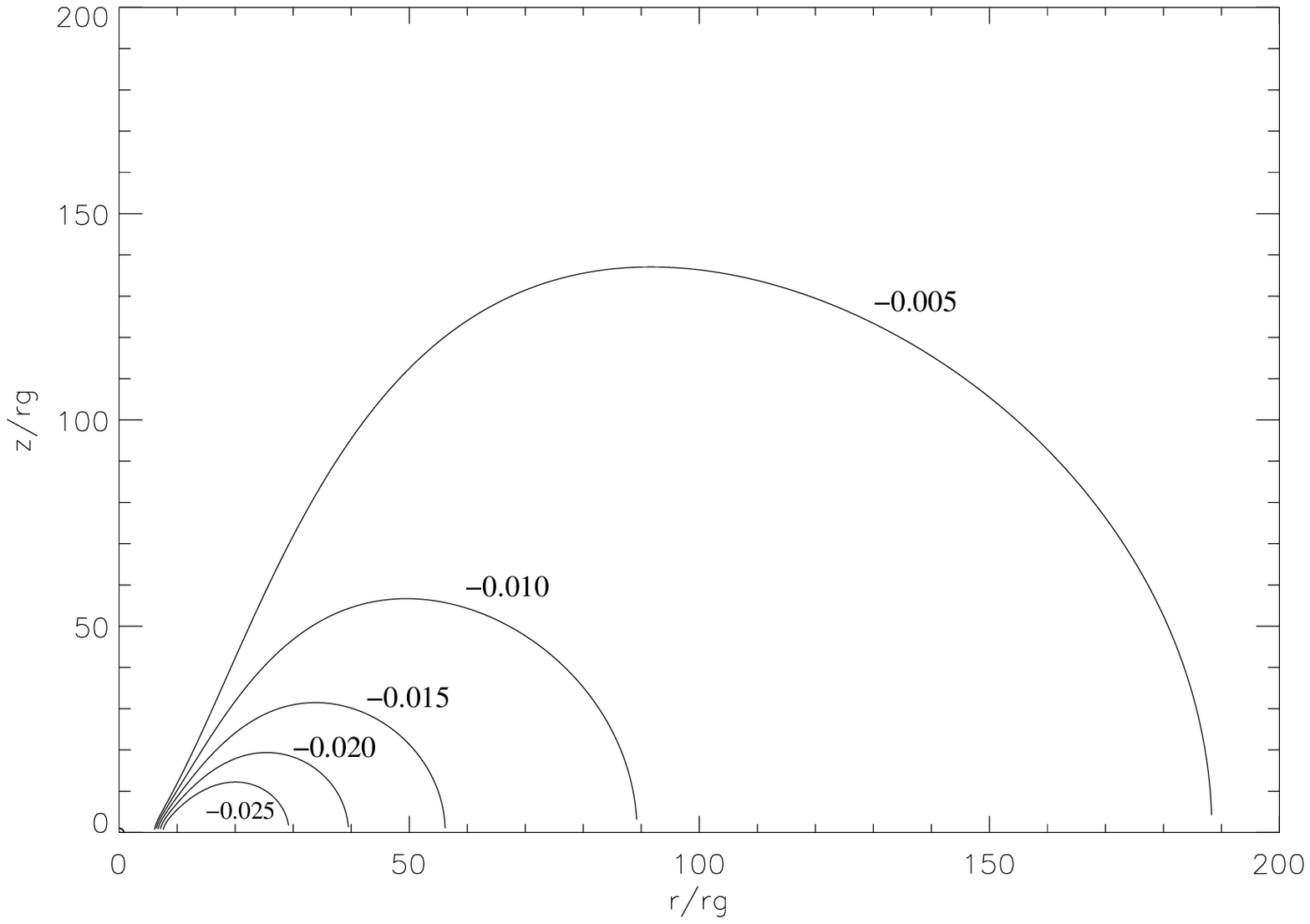}
\caption{Meridional cross-sections over one half the $r$--$z$ plane
for tori with constant specific angular momentum, $\ell={\rm const}$
(above), and a power--law specific angular momentum distribution $\ell
\propto r^{\alpha}$, $\alpha = 0.1$ (bottom). The filling factor
$\Phi_{0}$ is indicated in units of $c^{2}/2$.}
\end{figure}
Several groups have studied the dynamics and stability of tori with
varying distributions of specific angular momentum
\citep{df,fd1,mne,dm}. In summary, the conclusion has been that in a
torus with non--constant angular momentum, the runaway radial
instability is suppresssed.  We have performed calculations (described
below) in this context, and have used the value $\ell = 1.92 r_g c$
for a constant distribution.  This corresponds to the angular momentum
of a test particle in a Keplerian motion at $r=4.25r_g$. For the power
law distribution case we used
\begin{equation}
\ell(r)=\ell_K \left(\frac{r}{r_0}\right)^{\alpha},
\end{equation}
where
\begin{equation}
\ell_K= \left(r^3 \frac{\partial \Phi_e}{\partial r}\right)_{r=r_0}^{1/2}=
 \left[ \frac{GM_{BH}r^3}{(r-rg)^2} \right]_{r=r_0}^{1/2}.
\end{equation}
The value of the power index $\alpha$, can be super-keplerian
$\alpha>0.5$, keplerian $\alpha=0.5$ or sub--keplerian $\alpha < 0.5$.
To compare with the results obtained by \citet{fd1}, we have used a
sub--keplerian distribution with an index $\alpha= 0.1$. Finally the
normalizing factor for the radius is that which corresponds to the
Keplerian value $r_0=5.1r_g$.  We will henceforth define the centre of
the torus as the locus (a circle, really) of maximum density and
pressure in the disc.

\begin{figure}
\includegraphics[width=84mm]{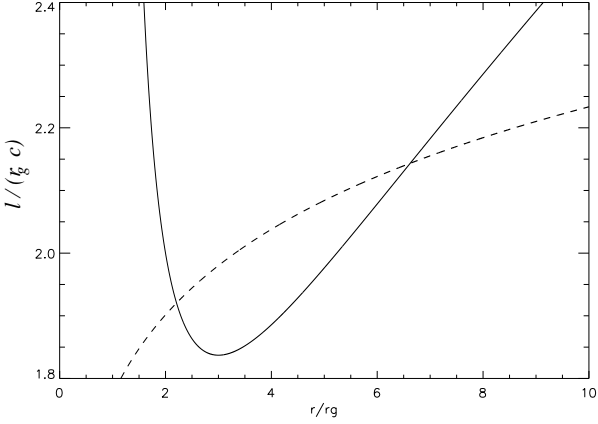}
\caption{Angular momentum distributions used for our models.  The
solid line represents a Keplerian angular momentum distribution for
test particles in circular orbits, while the dashed line shows the
non--constant power law distribution. For the case with uniform
angular momentum we used $\ell=1.92~r_{g} c$.}
\end{figure}

\subsection{Numerical method}\label{method}

We have used the SPH (Smooth Particle Hydrodynamics) numerical method
to perform our simulations \citep{mo}. Essentially this method works
by interpolating a desired function $A(r)$, with a set of given
quantities $A(r')$ throught a \emph{kernel} function $W(r,h)$, using
the following convolution integral:
\begin{equation}
A(r)=\int A(r')W(r-r')dr.
\end{equation}
The implementation of the code was made using the prescriptions
calculated by \citet{ml} when the system has azimuthal symmetry, using
the following \emph{kernel} function:
\begin{equation}
W(r,h)= \frac{\sigma}{h^{\nu}} \left\{ \begin{array}{l l}
         1-3(r/h)^{2}/2+3(r/h)^{3}/4 & \mbox{if $0
         \leq r/h \leq 1$;} \\ (2-r/h)^{3}/4 &
         \mbox{if $1 \leq r/h \leq 2$;} \\ 0 & \mbox{2 $\leq
         r/h$.}
         \end{array}
         \right.
\end{equation}
In the last equation, $h$ represents the smoothing length, and is
comparable to the typical separation between fluid elements, and $r$
is the radial coordinate ($h$ is essentially the spatial resolution of
the calculation).  As our calculations were performed in two
dimensions, $\nu=2$ and $\sigma=10/7\pi$.  We did not include the
viscosity of the gas and hence the Navier--Stokes equations of motion
take the form:
\begin{equation}
\frac{dv_{r}}{dt}=- \frac{1}{\rho}\frac{\partial P}{\partial r}
-\frac{GM_{BH}r}{R(R-r_g)^{2}}+r \Omega^2
+\left(\frac{dv_{r}}{dt}\right)_{art},
\end{equation}
\begin{equation}
\frac{dv_{z}}{dt}=- \frac{1}{\rho}\frac{\partial P}{\partial z}
-\frac{GM_{BH}z}{R(R-r_g)^{2}}+
\left(\frac{dv_{z}}{dt}\right)_{art},
\end{equation}
where again $R=\sqrt{r^2+z^2}$ is the distance to the central object.
The sub--index \emph{art} indicates the artificial viscosity terms,
used in the numerical calculations to account for the presence of
shocks and avoid particle interpenetration. We have additionally for
the conservation of angular momentum:
\begin{equation}
\frac{d \ell}{dt}=0.
\end{equation}
And finally, for the energy equation we have:
\begin{equation}
\frac{d \epsilon}{dt}=- \left(\frac{P}{\rho}\right)\nabla \cdot \vec{v}+
 \left( T \frac{ds}{dt} \right)_{art}
\end{equation}
where $\epsilon$ is the internal energy per unit mass. We have assumed
that we do not have losses of energy by external causes (i.e. $\Delta
Q=0$) and the only changes in $\epsilon$ arise from mechanical work
done on the system. In discretized form for numerical work, the
convolution integral given in equation~(8) becomes a sum over all the
fluid elements in the calculation.

For the artificial viscosity, we have used the prescription given by
\citet{ba} which reduces the artificial shearing stress during the
time evolution of the disc. The explicit forms for the acceleration
and the energy dissipation due to artificial viscosity for the $i-eth$
SPH particle are then:
\begin{equation}
\left(\frac{d \vec{v}}{d t}\right)_{i,art}=
- \sum_{j \neq i}m_{j} \Pi_{ij}\nabla_{i}W_{ij},
\end{equation}
and
\begin{equation}
\left(T \frac{ds}{dt}\right)_{i,art}= \frac{1}{2} \sum_{j \neq i}m_{j}
\Pi_{ij}(\vec{v_{i}}-\vec{v_{j}}) \cdot \nabla_{i}W_{ij}.
\end{equation}
where, again, $W$ is the \emph{kernel} function and $\Pi$ is defined
by \citep[see e.g.,][]{leerm}
\begin{equation}
\Pi_{ij}=\left(\frac{P_i}{\rho_i^2}+\frac{P_j}{\rho_j^2}\right)=
(-\alpha_b\mu_{ij}+\beta_b\mu_{ij}^2),
\end{equation}
\begin{equation}
\mu_{ij}= \left\{ \begin{array}{l l} \frac{(\vec{v}_i-\vec{v}_j) \cdot
(\vec{r}_i-\vec{r}_j)}
{h_{ij}(|\vec{r}_i-\vec{r}_j|^2/h_{ij}^2)+\eta^2}\frac{(f_i+f_j)}{2c_{ij}}
& \mbox{if $(\vec{v}_i-\vec{v}_j) \cdot (\vec{r}_i-\vec{r}_j) < 0$;}
\\ 0 & \mbox{if $(\vec{v}_i-\vec{v}_j) \cdot (\vec{r}_i-\vec{r}_j)
\geq 0$;}
         \end{array}
         \right.
\end{equation}
In this last equation the function $f_i$ is defined by:
\begin{equation}
f_i=\frac{|\nabla \cdot \vec{v}|_i}
{|\nabla \cdot \vec{v}|_i+|\nabla \times \vec{v}|_i + \eta'c_i/h_i}.
\end{equation}
The factor $\eta' \simeq 10^{-4}$ in the denominator prevents
numerical divergences, $c_i$ represents the sound speed and $h_i$ is
the lenght scale for the $i-eth$ particle.
$\alpha_b=\beta_b=\gamma/2$ are constants of order unity and $\gamma$
is the polytropic index from the equation of state. This form of the
artificial viscosity supresses the shearing stresses when the
compression in the fluid is low and the vorticity is high $|\nabla
\cdot \vec{v}| \ll |\nabla \times \vec{v}|$, but remains in effect if
the compression dominates in the flow $|\nabla \cdot \vec{v}| \gg
|\nabla \times \vec{v}|$.

\subsection{Initial conditions}\label{ICs}

Our initial conditions are generated by fixing the mass of the central
object, the equation of state, with $\gamma=4/3$ and the constant $K$,
the filling factor for the potential, $\Phi_{0}$ and the distribution
of angular momentum $\ell(r)$.  We then generate the fluid elements
which will model the torus using a Monte Carlo random number
generator. The particles that fall inside the torus density profile
will be the fluid elements of the simulation.  These fluid elements
are then relaxed for several dynamical times with an artificial
damping term included in the equations of motion in order to obtain
the a distribution which is as close to equilibrium as possible.  We
are able to build tori of different sizes by simply varying the
filling factor $\Phi_0$ as was shown in Figure~1.

Our initial configurations are always non--accreting tori, confined
inside their Roche lobe, i.e., $r_{in} > r_{L_1}$ where $r_{in}$ is
the distance from the central object to the inner edge of the torus
and $r_{L_1}$ is the radius of the inner Lagrange point, $L_{1}$.
When we focus on the pure oscillation dynamics of the disc, we also
wish to avoid accretion. In those cases, this is accomplished by
maintaining the mass of the black hole constant (even if some
particles overflow the Roche lobe as a result of the applied
perturbation).

Our code produces different snapshots of the main hydrodynamical
variables of the system like the position of the centre of the disc,
and also the maximum and mean densities and the potential, kinetic and
internal energies as functions of time.  This data then is analysed by
calculating the corresponding Fourier transforms and looking for
characteristic frequencies.

\subsection{Testing the code}\label{tests}

To test our code, we performed simulations reproducing previously
published resutls (\citet{acn}, \citet{dm}, \citet{akl}, \citet{fd1}
and \citet{df}) investigating the behavior of the runaway--radial
instability with changing distributions of angular momentum (in
systems without self--gravity).  We constructed tori that (i) almost
fill their effective potential wells, i.e., $r_{in} \simeq r_{L_1}$,
and (ii) have constant and non constant angular momentum
distributions.  We then introduced radial impulsive perturbations,
adding a velocity field to the initial distribution. This field is
similar to the solution proposed by \citet{mi} for spherical
relativistic accretion, and has the form
\begin{equation}
\begin{array}{l}
\vec{v}_r=- \eta \cdot \sqrt{r_{g}/R} \, \mathbf{\hat{r}}, \\
\vec{v}_z=- \eta \cdot \sqrt{r_{g}/R} \, \mathbf{\hat{z}},
\end{array}
\end{equation}
where the parameter $\eta \ll 1$ modulates the strength of the
perturbation.

We have found results that are in complete agreement with the previous
work available in the literature, that is, the runaway instability is
completly supressed when one considers a non--constant angular
momentum distribution within the torus.  This is shown in Figure~3,
where the mass of the disc is plotted as a function of time for a
calculation with a power law distribution of specific angular momentum
(solid line), and another with constant specific angular momentum
(dashed line). The former is clearly stable, while the latter is not,
and shows the characteristic runaway behavior. Once the instability is
triggered, the torus is destroyed in a few dynamical times.

\begin{figure}
\includegraphics[width=84mm,height=65mm]{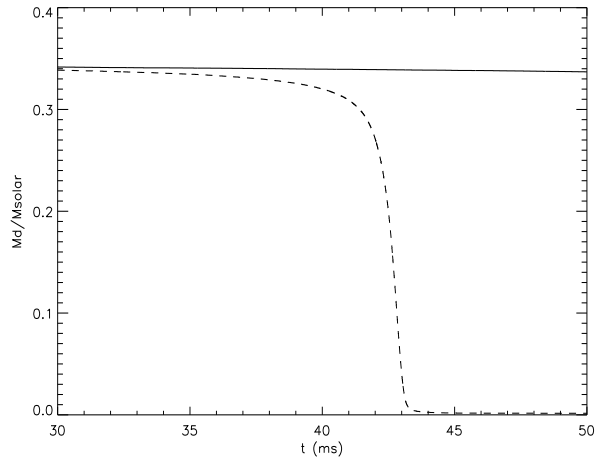}
\caption{Mass in the disc as a function of time for two test runs. The
dashed line corresponds to the model with constant angular momentum
distribution within the disc ($\ell={\rm cst}$), while the solid
line corresponds to the model with a power law angular momentum
distribution $\ell \propto r^{\alpha}$.}
\end{figure}

\section{Natural oscillations of the disc}\label{freeosc}

To study the free oscillations of the disc, we have performed
calculations where small, impulsive perturbations have been applied to
the velocity field, as in equation (19), only in the initial
conditions. This was done in particular to study how the thickness of
the disc affects its modes of oscillation. The space--time around the
black hole was kept constant during this set of runs, even if slight
overflow occurred through the inner Lagrange point.

\subsection{Epicyclic frequencies}\label{epi}

The results obtained were compared with the behaviour of a test
particle in a circular orbit around a central mass.  If we introduce a
small radial perturbation, the particle will show radial oscillations
at the epicyclic frequency $\kappa$. Our study begins with the
equation for radial motion of the particle:
\begin{equation}
\frac{d^2 r}{dt^2}=-\frac{d\Phi_{\rm eff}}{dr}.
\end{equation}
Introducing a perturbation of the form: $\Delta r=r-r_0$, and
calculating a first order Taylor expansion over $r$ for small
perturbations, we obtain
\begin{equation}
\frac{d^2 \Delta r}{dt^2}=-\left( \frac{d^2\Phi_{\rm eff}}{dr^2} \right)_0
\Delta r,
\end{equation}
for a fixed radius. This equation is simply that of a harmonic
oscillator, with epicyclic frequency $\kappa=(d^2\Phi_{\rm
eff}/dr^2)_0^{1/2}$.  For a constant angular momentum ($\ell={\rm
cst}$), we have:
\begin{equation}
\kappa (r)=\frac{1}{2 \pi}\sqrt{\frac{GM_{BH}(r-3r_g)}{r(r-r_g)^3}}.
\end{equation}
In Figure~4 we show the radial dependency of $\kappa(r)$. \\

\begin{figure}
\includegraphics[width=84mm]{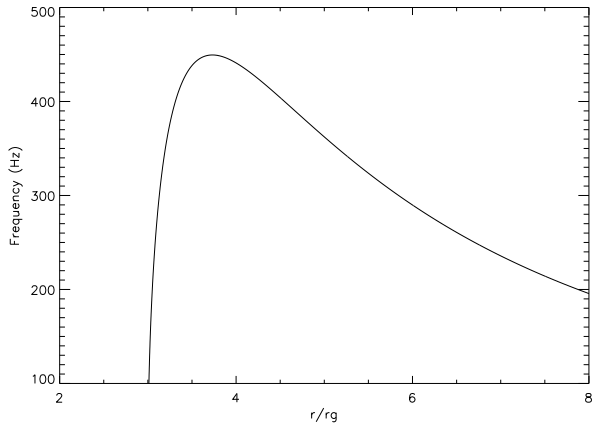}
\caption{Epicyclic frequency $\kappa$ as a function of radius. The
  curve is computed for a test particle orbiting a central object of
  2.5$M_{\odot}$. Compare with Figures~5 and 6.}
\end{figure}

\subsection{Results}\label{results1}

We show here results for tori of three different sizes: small ($r_d <
r_g$), intermediate ($r_d \sim r_g$) and large ($r_d > r_g$), both for
$\ell =$cst and $\ell(r)=r^{\alpha}$.  In each case we computed the
Fourier transform of the radial motions of the centre of the
disc. This always exhibited one prominent peak, at frequency
$\kappa^{*}$. In the limit of small tori, $\kappa^{*} \rightarrow
\kappa(r_{0})$, the value for test particle motion. As the size of the
torus increases, there is a shift to lower frequencies, and
$\kappa^{*}$ decreases. The Fourier transforms and the corresponding
values for the frequencies (along with the model parameters) are given
in Figures~5 and~6, and in Table~1. These numerical results agree with
the work reported by \citet{rymz} and by \citet{ryz}, where the
frequency shift has been computed analytically in the Schwarschild
metric (albeit with the aforementioned simplification of
height--integrated equations). We thus confirm the p--mode
interpretation of Rezzolla et al. (2003), in which \emph{the radial
epicyclic frequency at the position of the centre of the torus
represents the value at which the fundamental p--mode frequency tends
to in the limit of vanishing torus size.}

\begin{table*}
\begin{minipage}{126mm}
\caption{Models for discs of different size orbiting a 2.5M$_{\odot}$
black hole. We give the extent of the disc $r_d$, the disc to black
hole mass ratio (M$_d/$M$_{BH}$), the angular momentum value
($\ell_0/r_gc$) or distribution ($\ell \propto r^{\nu}$), the locus of
the point of maximum density ($R_{c}/r_g$), the filling factor which
gives us the size of the torus ($\Phi_0/(c^2/2)$), the eigenfrequency
$\kappa^{*}$ at which the point of maximum density oscillates, and the
number of SPH particles we used in the simulation ($N$).}
\centering
\begin{tabular}{l c c c c c c c}
\hline
Model    & size ($r_d$) & M$_d/$M$_{BH}$       & $\ell_0/r_gc$     & $R_{c}/r_g$  & $\Phi_0/(c^2/2)$  & $\kappa^{*}$ (Hz)  &  N \\
\hline
(a) &  $r_d < r_g$     & $3.0 \times 10^{-8}$   &  1.9197             &  4.25        & -0.1183  & 425   & 2337  \\
(b) &  $r_d \sim r_g$  & $5.9 \times 10^{-3}$   &  1.9197             &  4.25        & -0.1060  & 355   & 2886  \\
(c) &  $r_d > r_g$     & 0.14                   &  1.9197             &  4.25        & -0.0841  & 290   & 4082  \\
\hline
(d) &  $r_d < r_g$     & $1.6 \times 10^{-8}$   & $\propto r^{\nu}$ &  5.10        & -0.0749  &  350  & 2317 \\
(e) &  $r_d \sim r_g$  & $2.0 \times 10^{-3}$   & $\propto r^{\nu}$ &  5.10        & -0.0678  &  300  & 1866 \\
(f) &  $r_d > r_g$     & 0.3                    & $\propto r^{\nu}$ &  5.10        & -0.0595  &  250  & 1630 \\
\hline
\end{tabular}
\end{minipage}
\end{table*}

\begin{figure*}
\begin{minipage}{126mm}
\includegraphics[width=65mm]{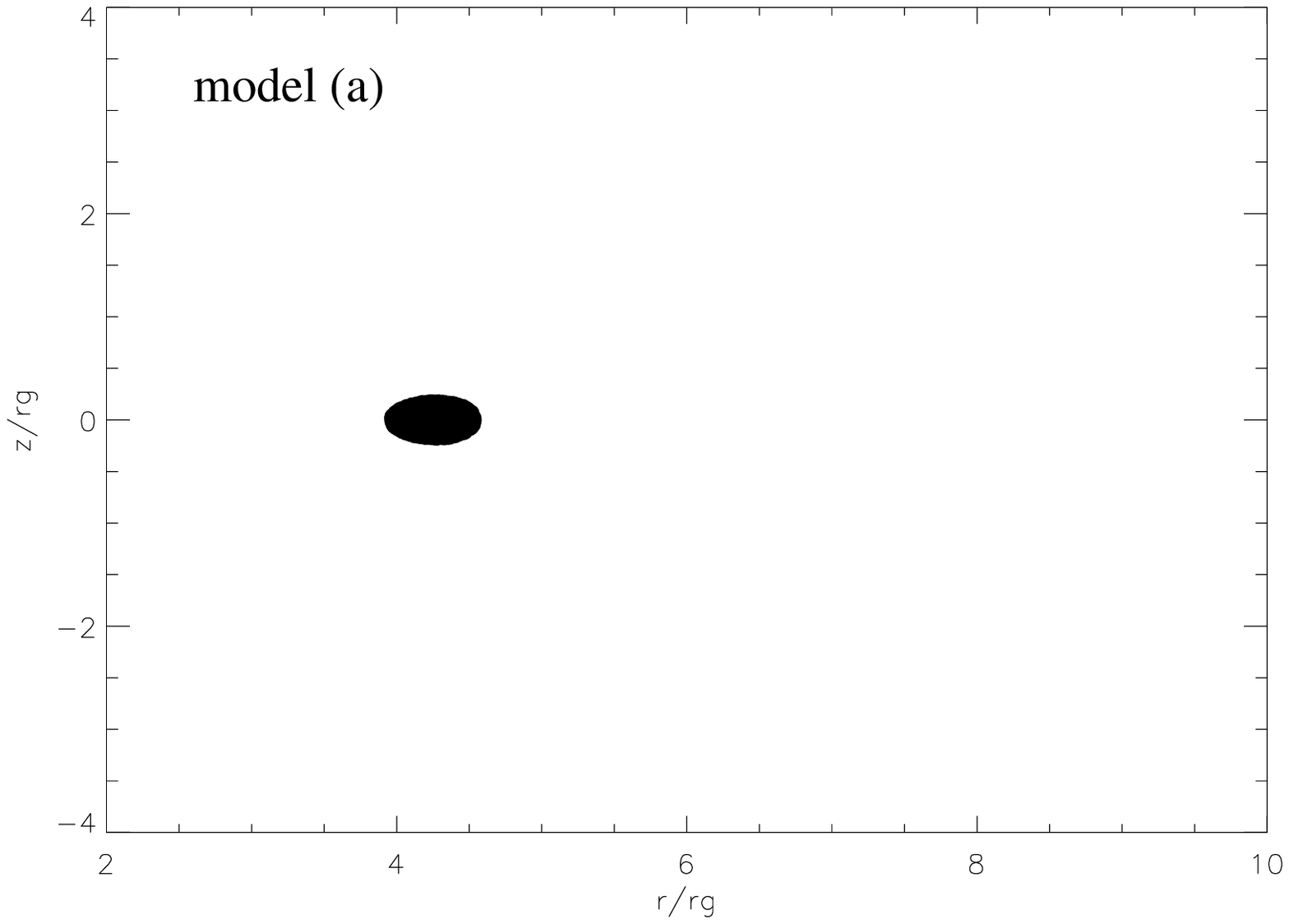}
\hspace{2mm}
\includegraphics[width=65mm]{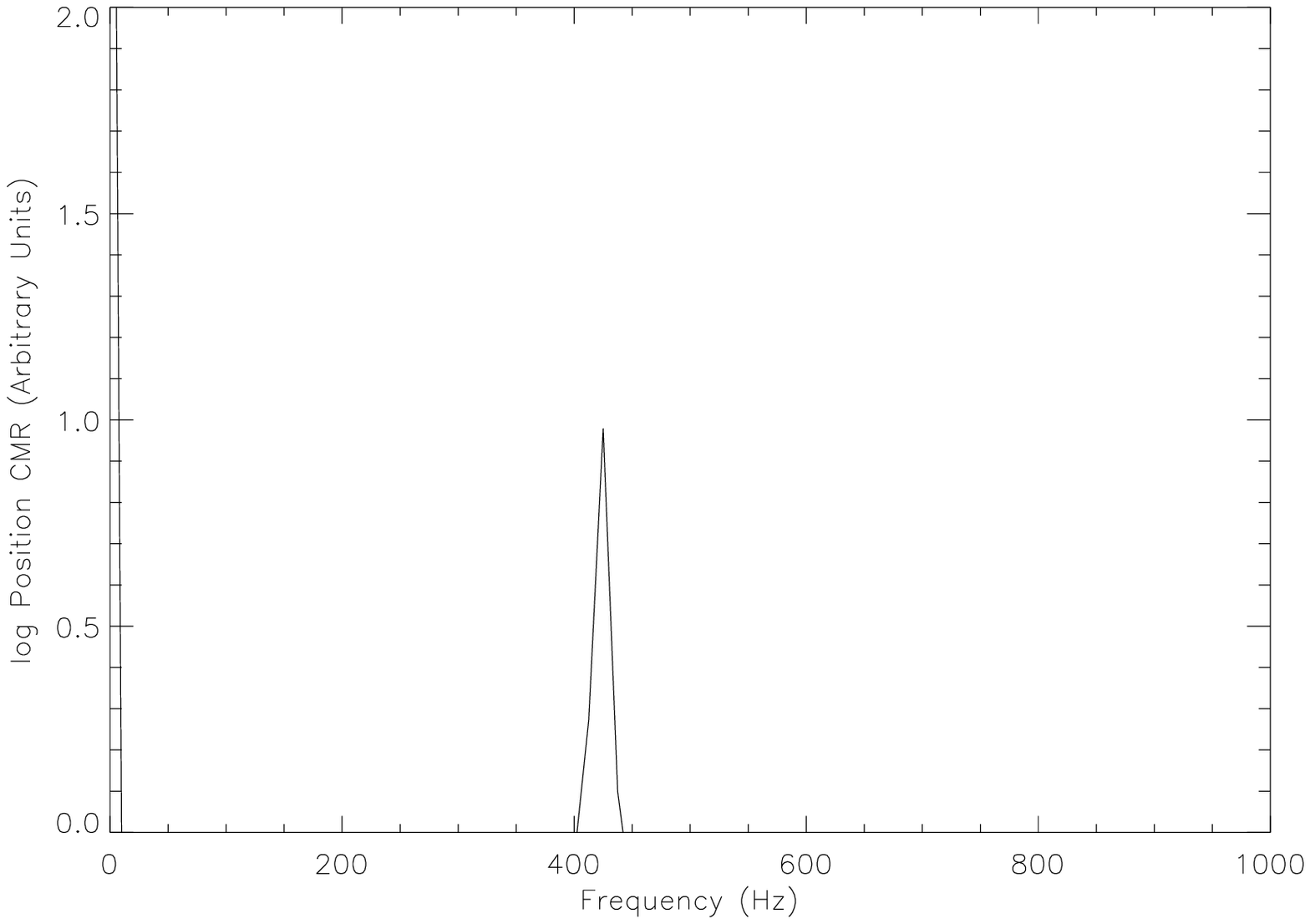} \\
\includegraphics[width=65mm]{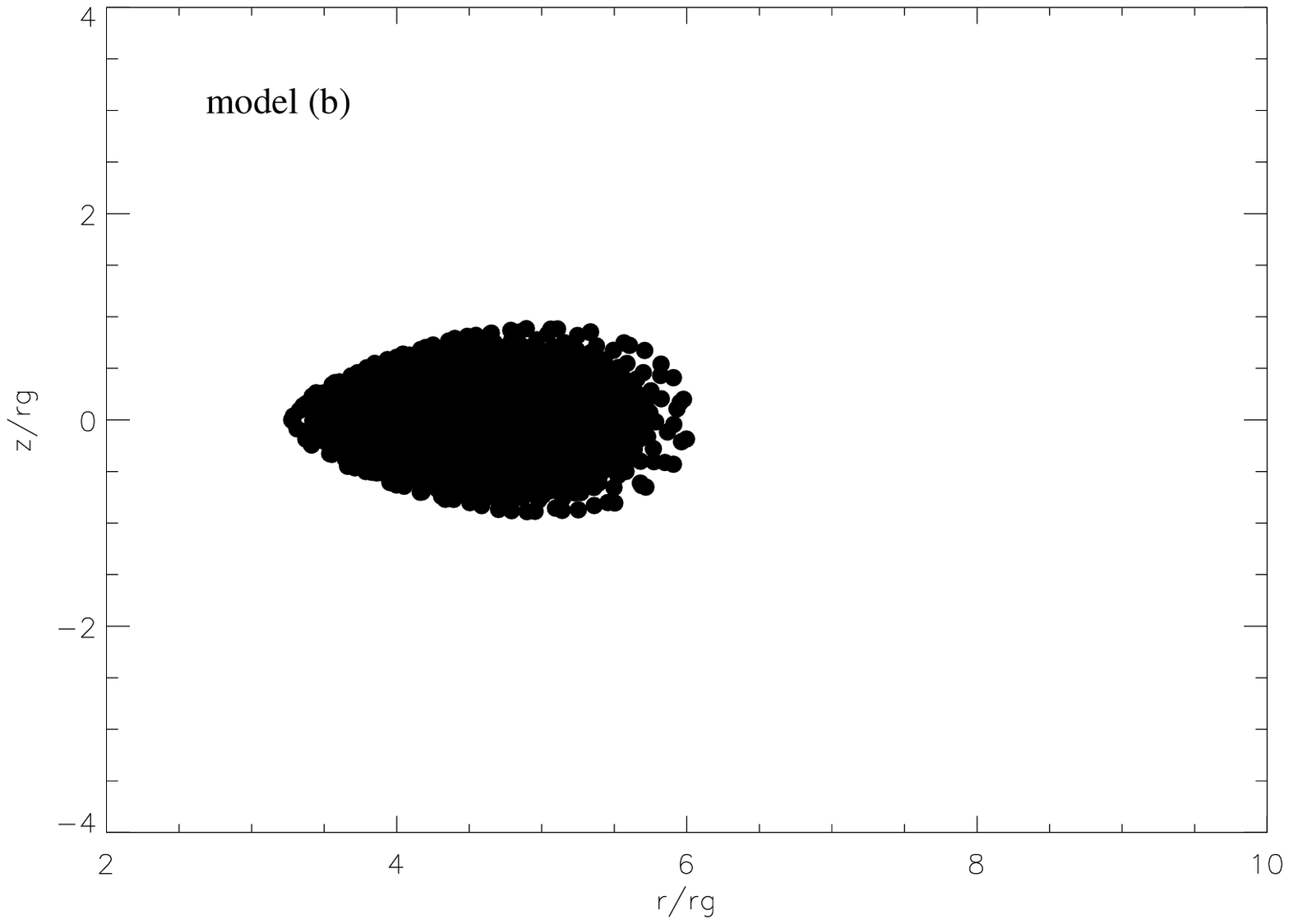}
\hspace{2mm}
\includegraphics[width=65mm]{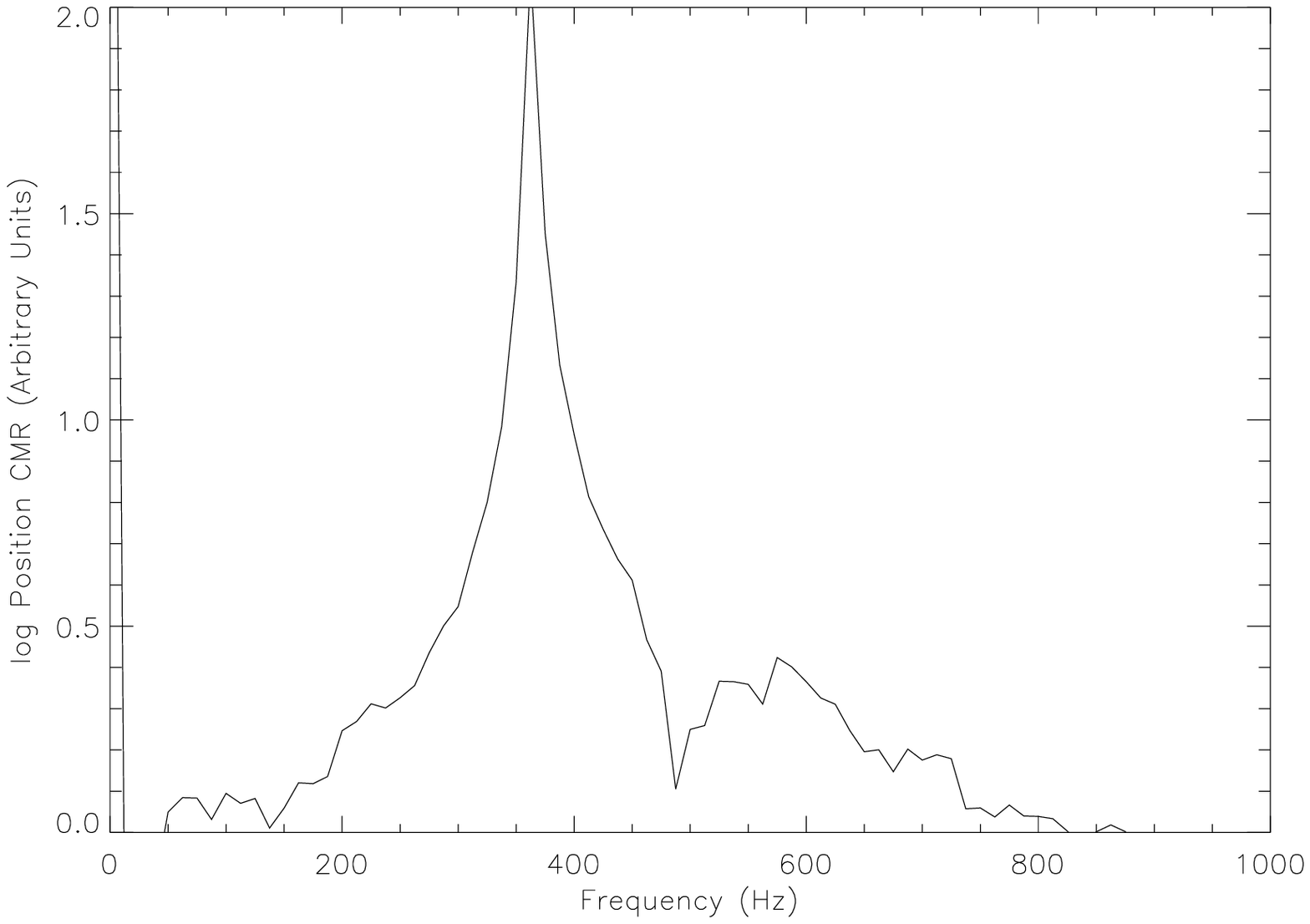} \\
\includegraphics[width=65mm]{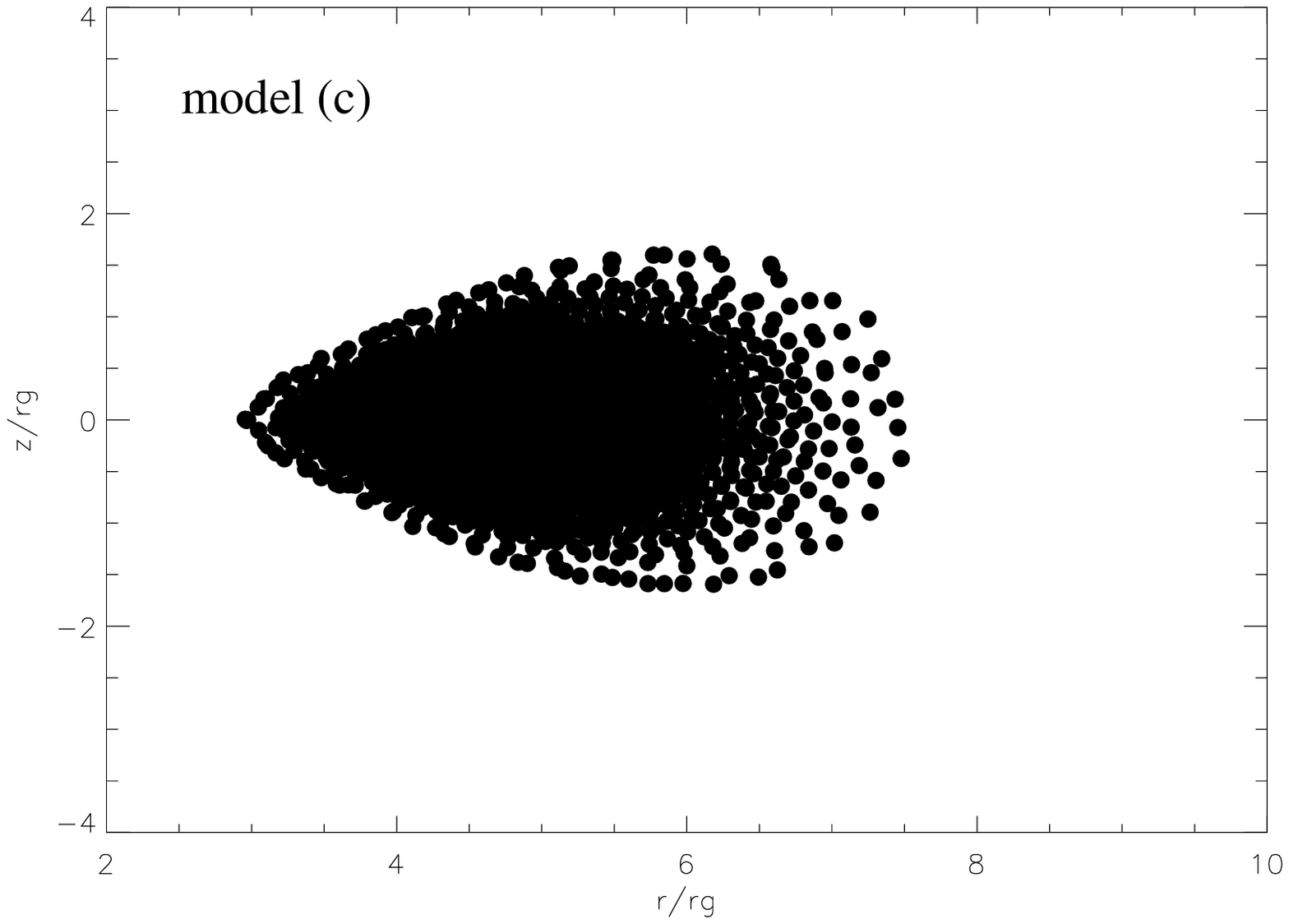}
\hspace{2mm}
\includegraphics[width=65mm]{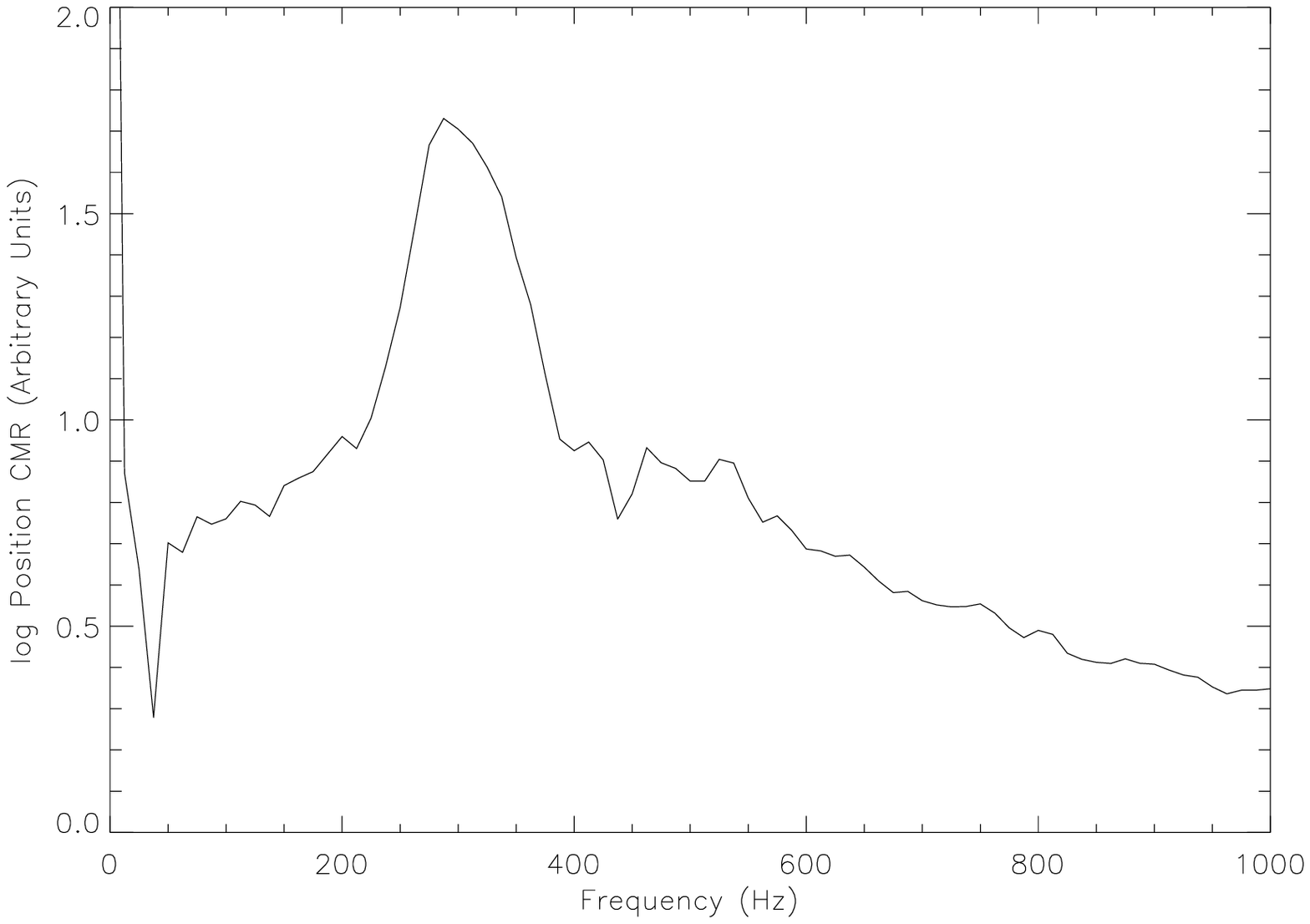} \\
\caption{Meridional cross sections for a small, intermediate and large
torus (left panels), for a constant angular momentum distribution
($\ell={\rm cst}$), and Fourier transforms of the radial oscillations of
the centre of the torus for each case (hereafter represented by the letters CMR in the figures)
in the right pannels.
The discs orbit
a 2.5M$_{\odot}$ black hole.  Note how the eigenfrequency $\kappa^{*}$
is progressively shifted to lower values as the size of the torus
increases and compare with the epicyclic eigenfrequency for a test
particle shown in Figure~4. For all the unperturbed tori shown here,
the centre is located at $r=4.25r_g$.}
\end{minipage}
\end{figure*}

\begin{figure*}
\begin{minipage}{126mm}
\includegraphics[width=65mm]{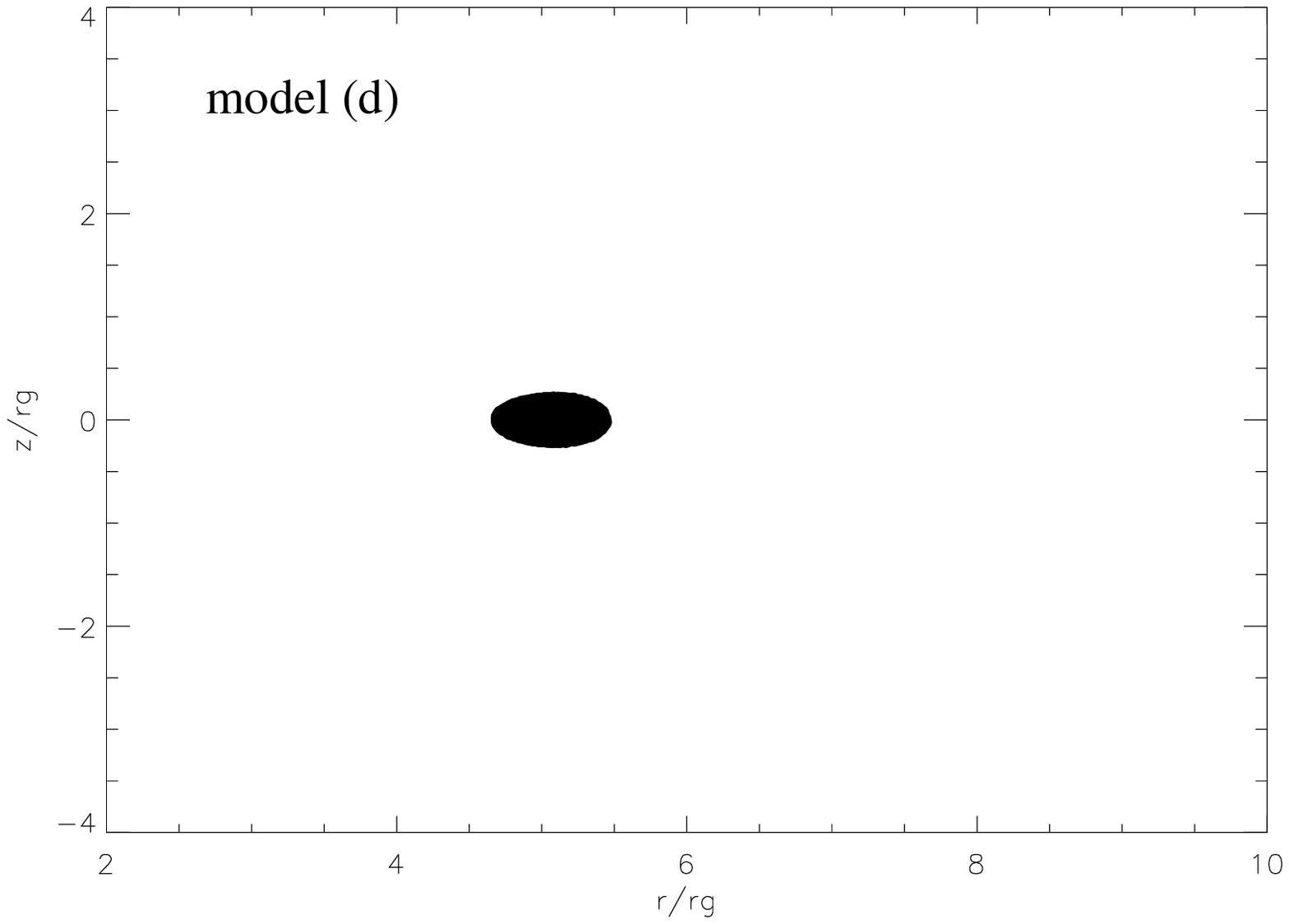}
\hspace{2mm}
\includegraphics[width=65mm]{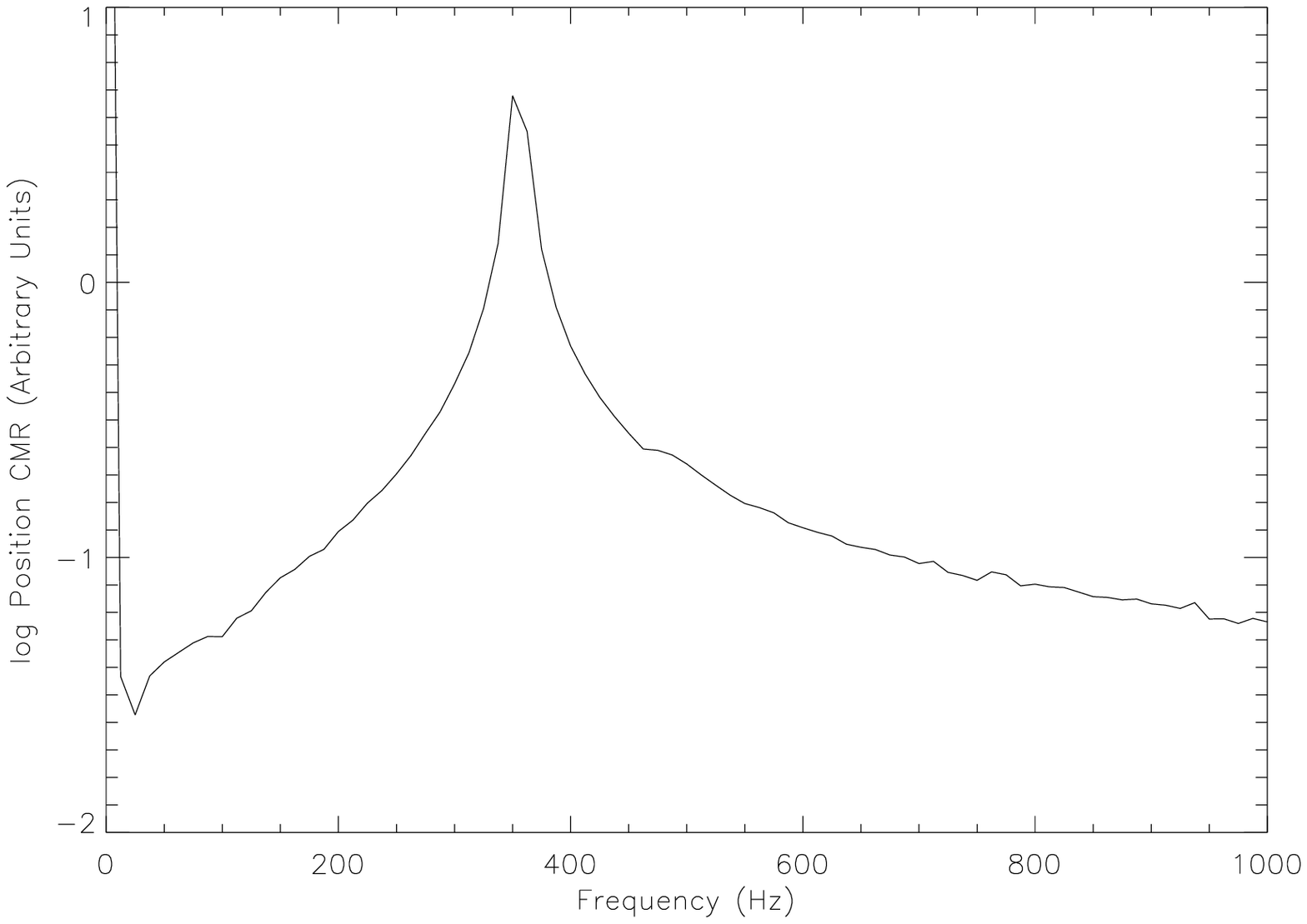} \\
\includegraphics[width=65mm]{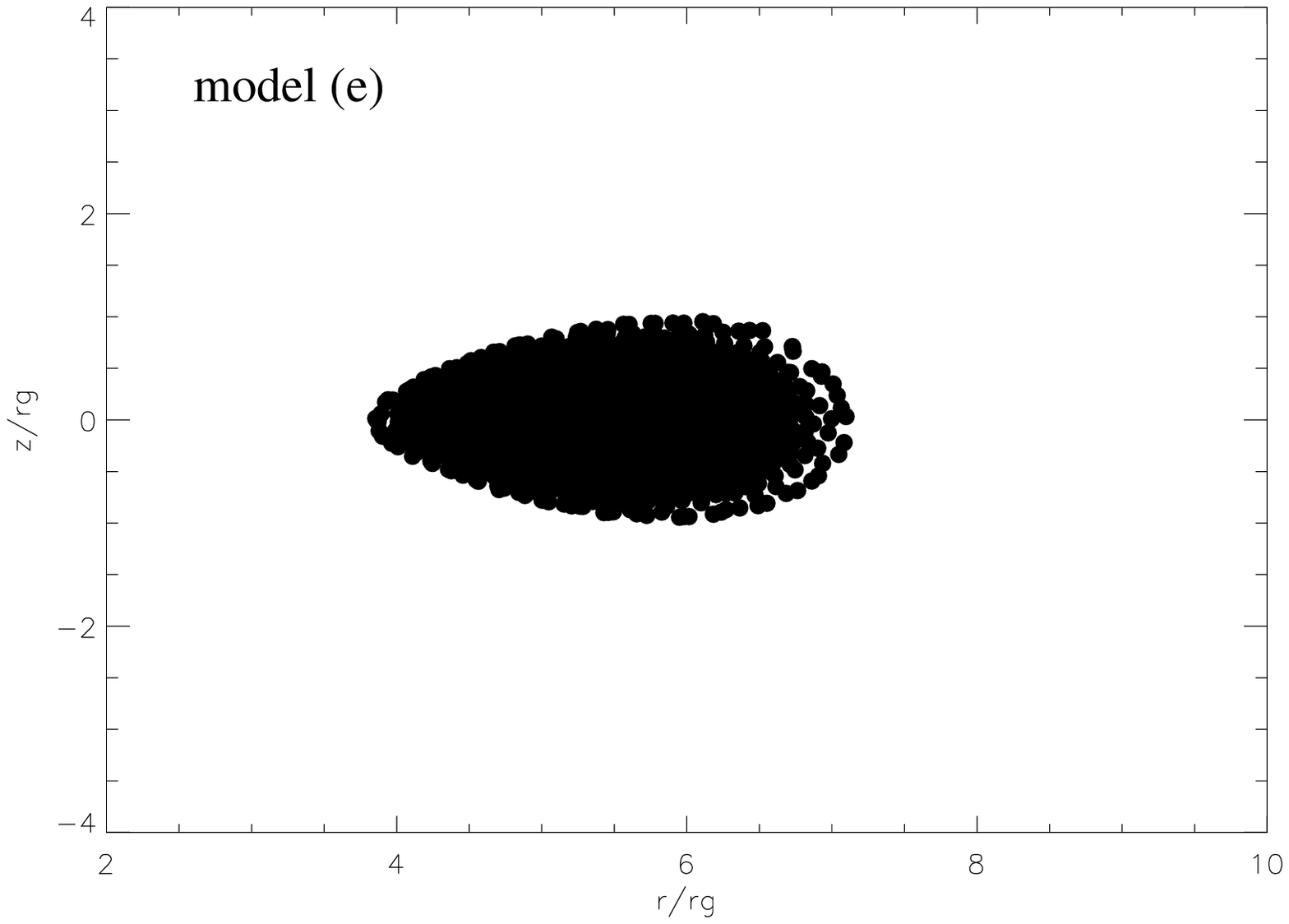}
\hspace{2mm}
\includegraphics[width=65mm]{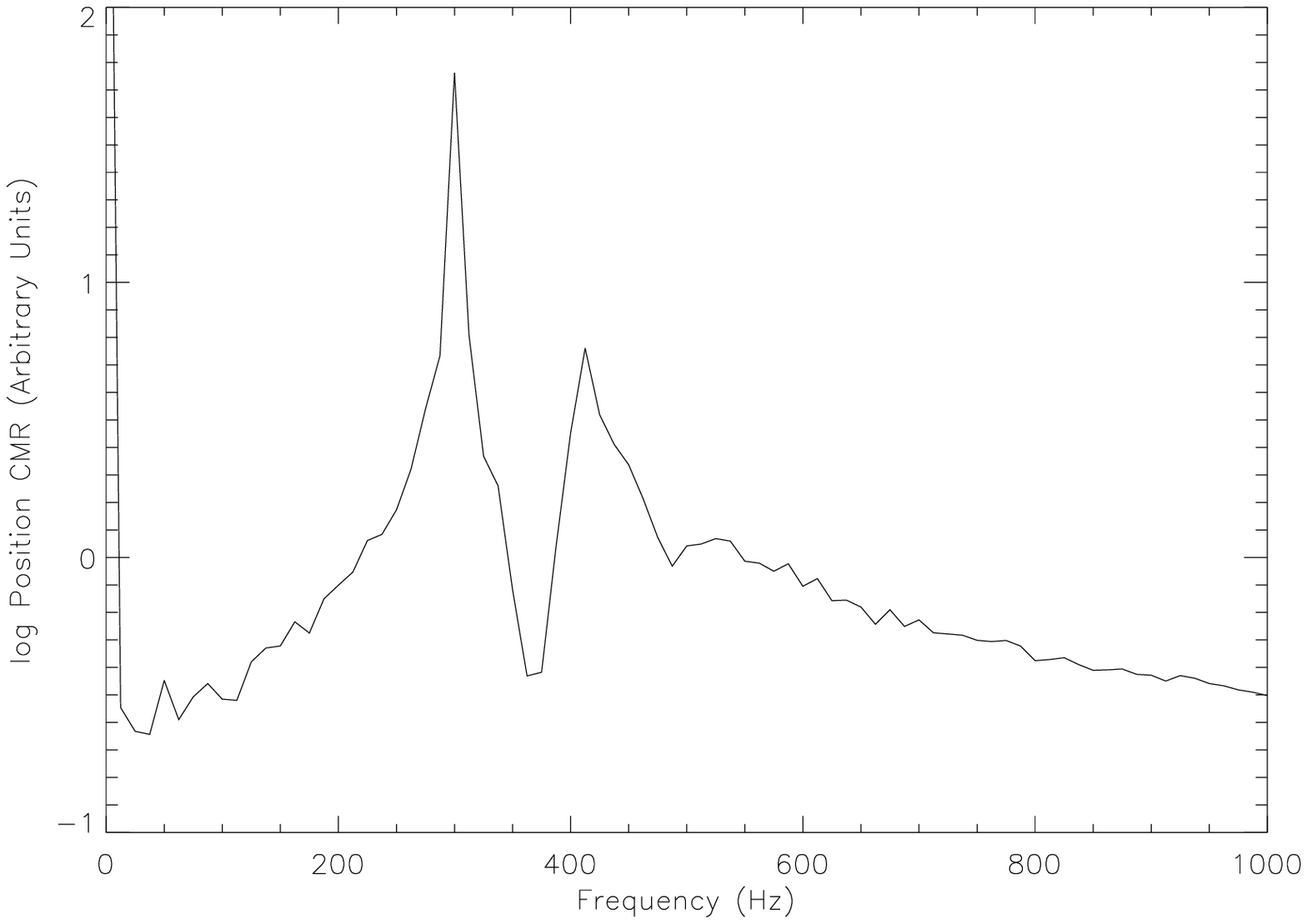} \\
\includegraphics[width=65mm]{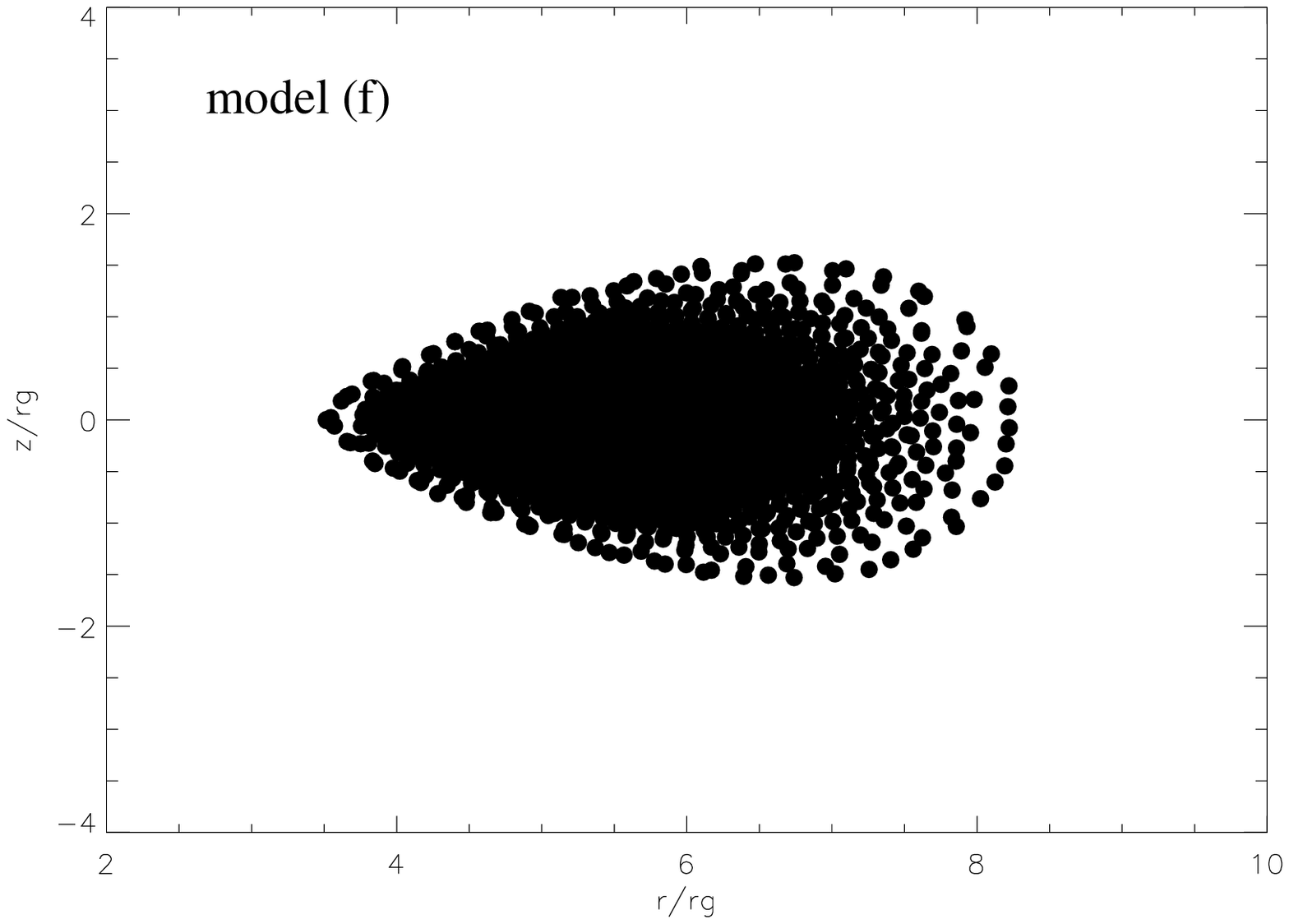}
\hspace{2mm}
\includegraphics[width=65mm]{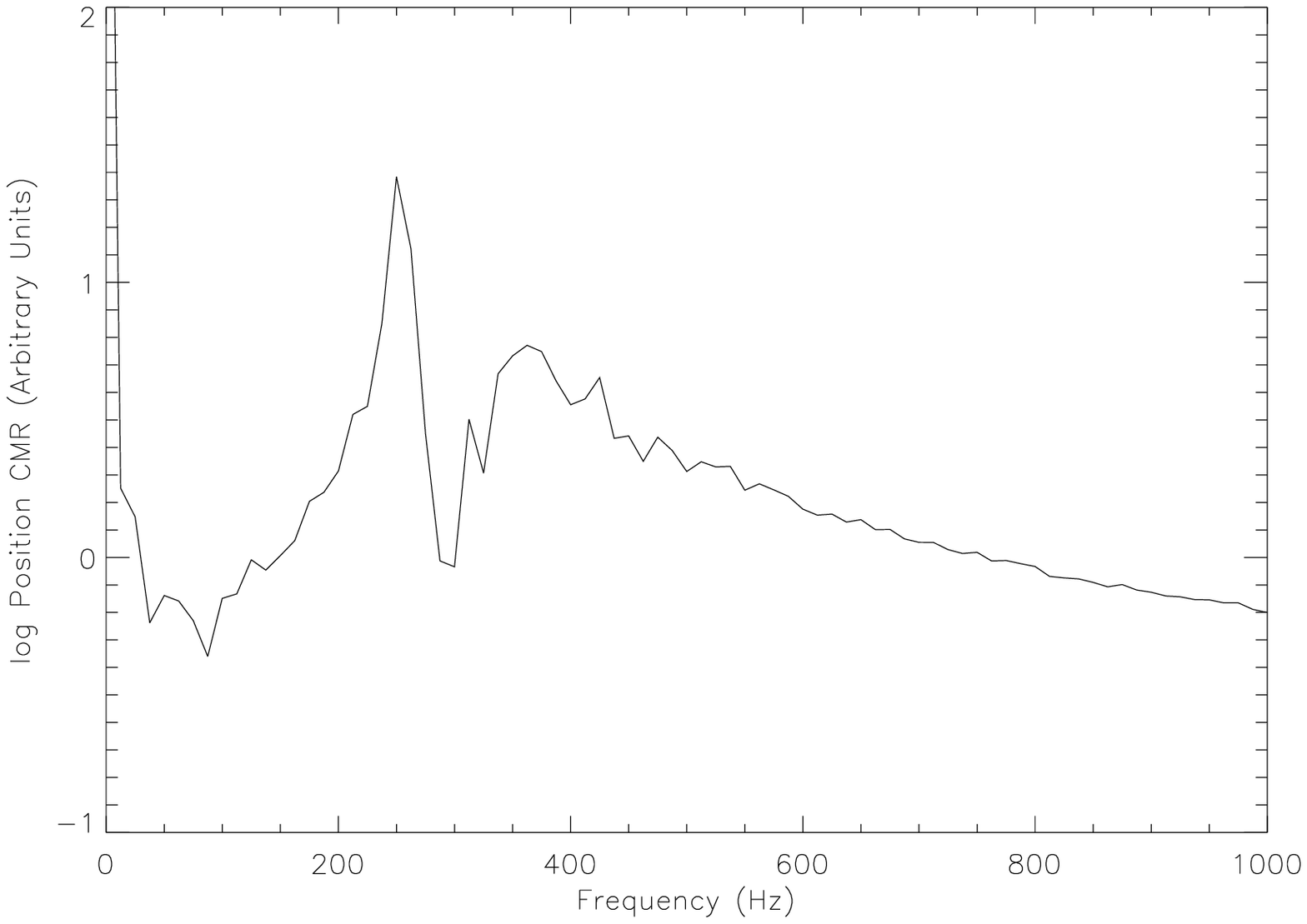} \\
\caption{Same as Figure~5, but for a power--law angular momentum
distribution ($\ell \propto r^{\alpha})$.  For all the unperturbed
tori shown here, the centre is located at $r=5.1r_g$.}
\end{minipage}
\end{figure*}

\section{Localized periodic perturbations and global response}\label{forcedosc}

\subsection{A localized external perturbation}\label{locpert}


Having identified the strongest response in the radial oscillations as
a modified epicyclic frequency, or p--mode, we proceeded to apply
localized, periodic perturbations to thick tori (as in Paper~I). We
believe it is reasonable to assume that the perturbation will be
stronger in the inner regions of the disc, if it arises from the
central object, and periodic, if it is related to its spin period
(clearly identifiable in neutron star systems) or some other mechanism
repeating at intervals $\Delta t=1/ \nu_{per}$.  This perturbation can
be related to a wide variety of phenomena coming from the central
object or from the material that conforms the disc.  A specific
example could be the emission of gravitational radiation from the
black hole when a clump of material is accreted and breaks the axial
symmetry in a cuasi--periodic fashion.  This could produce a time
variation in the mass quadrupole of the disc itself
\citep{zrf}. Another possibility in this context was proposed by
\citet{vp01}, and is related to the connection between the magnetic
field of the disc and the gravitational radiation produced by the
black hole. Disc oscillations can be related to the formation of the
disc, or to its intrinsic properties. Simulations by \citet{ica} have
shown that variablity and oscillations can be triggerred by viscous
processes. \citet{kato01} has suggested that energy flux within the
disc through viscous and angular momentum processes can alter the
equilibrium state, producing oscillations.  Finally, in highly
dynamical situations, the disc will naturally lack complete azimuthal
symmetry (e.g., in compact object mergers) and the formation of clumps
inside the disc could lead to small oscillations.

We have modeled perturbations in the context of non-constant angular
momentum discs in the same way that we did in Paper~I, introducing an
additional acceleration in the equations of motion, given by:
\begin{equation}
a_{pert}=-\eta a_g \cdot \exp \left(\frac{r_0-r}{\delta r}\right)
\cdot \sin(2 \pi \nu_{per} t).
\end{equation}
Here $a_g$ is the acceleration due to gravity, $r_0$ is the outer edge
of the torus and $\eta \ll 1$ is a parameter that modulates the
strength of the perturbation.  The exponential term decays on a scale
$\delta r \simeq R$, the radial extension of the disc, thus
reproducing the desired behaviour for the perturbative force, which
will be strong near the inner radius and weak in the outer regions.
This acceleration induces radial oscillations, which can be Fourier
analyzed to extract the main frequencies in a similar fashion as was
done by \citet{lak}.

\subsection{Results}\label{results2}

We performed simulations using constant and non--constant angular
momentum distributions within the disc. For the constant distribution,
we find, as described in Paper~I for a few particular cases, that the
localized perturbation induces global oscillations at various
characteristic frequencies. The most prominent (besides that ocurring
at the perturbing frequency $\nu_{per}$) is at $\nu_{1}=\kappa^{*}$, the
modified epicyclic frequency associated with the radial motions of the
centre of the disc. A second broad peak with about ten times less
power at $\nu_{2}$ is in a 3:2 relation with the former (see
Figure~7). This latter feature is barely made out in the calculations
where an impulsive perturbation was applied to the whole disc at
$t=0$. 

For a power law distribution of angular momentum we find that the disc
responds somewhat differently to the localized perturbation. A typical
power spectrum for the radial motions of the centre of the disc is
shown in Figure~8 (when $\nu_{per}=100$~Hz). Two strong narrow peaks
are evident at $\nu_{per}$ and $2\nu_{per}$, despite the fact that the
perturbation is a pure sinusoid. Furthermore, a weak subharmonic at
$\nu_{per}/2$ is also present. This is clear evidence of non--linear
coupling between various modes, and is reminiscent of analogous
results for forced oscillations in slender tori previously reported in
the context of kHz QPOs in LMXBs \citep{lak}. Note that the overtone
is absent in the power spectrum in Figure~7 (when the angular momentum
is constant in the disc). As for the case with constant angular
momentum, two more peaks, at $\nu_1=\kappa^{*}=250$~Hz, and $\nu_{2}
\approx 1.5 \nu_{1}$ can be seen in the spectrum. To the limit of our
resolution, they are in a 3:2 correspondence. Thus it appears that the
angular momentum distribution is somehow responsible for a certain
mode coupling which allows higher harmonics and subharmonics of the
perturbation to appear in the power spectrum.  Note also that when an
impulsive perturbation was used in a thick torus with a power law
distribution of angular momentum, secondary features, consistent with
being in a 3:2 relation with the stronger ones, are apparent in the
spectrum once the torus becomes relatively thick(see Figure~6, middle
and bottom right panels). Apparently the periodic nature of the
forcing also has a direct influence on the magnitude of the response
of the system. We elaborate further on this point below in
\S~\ref{discussion}.

\begin{figure}
\includegraphics[width=84mm]{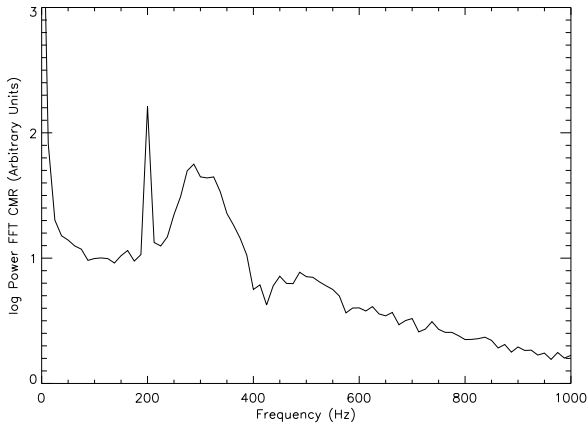}
\caption{Fourier power spectrum for the radial oscillations of the
centre of a torus (CMR) with a constant angular momentum distribution, when
perturbed periodically.  Three main peaks are apparent. From left to
right, the first one is the oscillation indcued at
$\nu_{per}=200$~Hz, the second is the modified epicyclic frequency
(p--mode) at $\nu_{1}=\kappa^{*}=300$~Hz, and the third is in a 3:2
relation with $\kappa{^*}$, at frequency $\nu_{2}=450$~Hz.}
\end{figure}

\begin{figure}
\includegraphics[width=84mm]{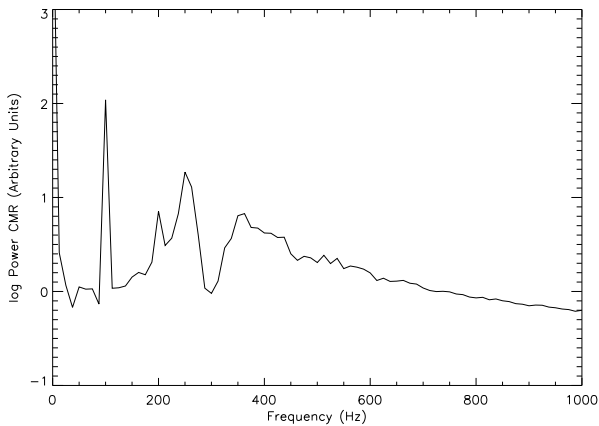}
\caption{Fourier power spectrum for the radial oscillations of the
centre of a torus (CMR) with a power law distribution of angular momentum,
$\ell(r) \propto r^{\alpha}$, when perturbed periodically. Several peaks are
apparent. The strongest one is at the perturbation frequency,
$\nu_{per}=100$~Hz. A harmonic at 200~Hz and a weak subharmonic at
50~Hz are also visible. The fundamental p--mode is at
$\nu_{1}=\kappa^{*}=250$~Hz, and the weaker first overtone at
$\nu_{2}=360$~Hz. The latter two are approximately in a 3:2 ratio.}
\end{figure}

\section{DISCUSSION}\label{discussion}

We have performed a numerical study of the global oscillations of
geometrically thick fluid tori in axisymmetry. We have neglected
self--gravity, but used a pseudo--potential to mimic the effects of
strong gravity for the potential produced by the central mass. After
building initial conditions in hydrostatic equilibrium with a
prescribed rotation law (given through the distribution of specific
angular momentum within the disc), global impulsive and localized
periodic perturbations have been applied, and the disc's response has
been measured through the time variation of certain quantities, and
their corresponding Fourier transforms. An ideal gas equation of state
has been used throughout. 

As the size of the torus is increased by filling up the effective
potential well to a greater degree (see Figures~1, 5 and 6), we find,
by applying impulsive kicks, that the fundamental global mode of
radial oscillations monotonically decreases from the analytic
expression for the local radial epicyclic frequency at the position of
the centre of the torus. This effect is present independently of the
assumed distribution of angular momentum (be it a constant or a power
law). That is, a slender torus tends to oscillate as a free particle
would, and a thicker one does so at lower frequencies. This effect is
simply due to the increased size of the resonant cavity, and the
density stratification within it which affects the propagation of
sound waves. It was found analytically in calculations performed in
the Schwarzschild metric with height--integrated tori for
inertial--acoustic modes (p--modes) recently \citep{rymz,ryz}, and we
confirm it here for the more general case of a full torus perturbed
slightly from equilibrium.


Upon application of a localized and periodic external perturbation,
however, the torus responds in a manner which does depend on the
distribution of angular momentum. As previously reported in the
context of LMXBs (Paper~I), for $\ell$=cst the fundamental radial mode
at frequency $\kappa^{*}$ and the first overtone at $1.5\kappa^{*}$ are
excited (although the overtone is noticeably weaker than the
fundamental).

For a power law distribution of angular momentum (specifically we used
$\ell(r) \propto r^{0.1}$), the fundamental at $\kappa^{*}$ was always
present, as was the first overtone at $\approx1.5\kappa^{*}$.
Furthermore, a harmonic of the perturbation itself, at $2\nu_{per}$,
and a subharmonic at $\nu_{per}/2$ always appeared. We take this
latter fact to indicate the presence of non--linear mode coupling,
since the applied perturbation was purely sinusoidal and contained
only a single frequency.

Trial calculations in both cases revealed that varying the
perturbation frequency $\nu_{per}$ by a factor of a few had no
discernible effect on the results (i.e., this is not the result of a
coincidence of fundamental frequencies and overtones with the
particular choice for the perturbation frequency).

Similar results concerning the relative amplitudes of various modes
when $\ell(r)$ is varied can actually be extracted from the study of
Rezzolla and collaborators. In particular, Figure~3 in \citep{ryz}
shows the response (in the density) of a torus with constant angular
momentum when given an impulsive perturbation. The first overtone is
weaker than the fundamental by a factor of $\simeq 30$. However,
results for tori with power--law distributions in $\ell(r)$ show a
much larger contrast, with the fundamental and its harmonics
dominating the frequency spectrum, and p--mode overtones being much
weaker (about three orders of magnitude, see Figure~4 in
\citet{zfrm}). To complicate matters, it would appear that the way in
which the pertrubation is applied (e.g., only in the density vs. only
in the velocity, or both) has an important effect on the amplitude of
the response (see Figure~5 in \citet{zfrm}). In calculating the
eigenfunctions for the height--integrated problem, Rezzolla and
collaborators have found that these become vanishingly small near the
inner edge of the torus in certain cases, thus making these modes
harder to excite. In all cases, and as one would expect, they report
that the most efficient way to excite oscillations is to use a
perturbation which matches the corresponding eigenfunction (i.e., this
is when the strongest response occurs).  We have only attempted one
type of pertrubation here, given in equation~(25), and so have not
explored this effect. The vanishing of the eigenfunction may explain
formally why certain oscillations are suppressed, or equivalently, why
they are so hard to excite with arbitrary perturbations.

In the context of QPOs in LMXBs, producing variability in the X--ray
lightcurve from the oscillations described here is beyond the scope of
this work \citep[see e.g.,][for how this may actually come
about]{schnittman05}. However, if it is actually related to the QPOs,
one could argue that the presence of definite frequencies, and their
relation to one another, may serve as an indicator of the distribution
of angular momentum within the disc, or of its relative thickness.

We believe the study presented here captures the essentials of the
simple system at hand, namely, one with pure hydrodynamics in the
field of a central object, neglecting self gravity and assuming
azimuthal symmetry. Clearly in a real astrophysical situation these
conditions will not be strictly met. Magnetic fields are likely to
play an important role and make the system more complex by introducing
additional possible modes, as well as coupling to the
hydrodynamics. Even without magnetic fields, the tori described here
will not be isolated in a vaccuum, but surrounded by a gaseous
envelope, the rest of the disk. Mode coupling and/or leakage at the
interface is likely to affect the oscillatory behaviour of the system
and is a consideration that deserves further study \citep[for a
preliminary study addressing these matters see][]{fragile05}.

\section*{ACKNOWLEDGMENTS}

We gratefully acknowledge helpful discussions with M.~A. Abramowicz,
W. Klu\'{z}niak and L. Rezzolla. This work was supported in part by
CONACyT (36632E).

\bsp  

\label{lastpage}


\begin{thebibliography}{}

\bibitem[\protect\citeauthoryear{Abramowicz, Calvani \& Nobili}{1983}]{acn}
   Abramowicz M. A., Calvani M., Nobili L. 1983, ApJ, 242, 772

\bibitem[\protect\citeauthoryear{Abramowicz, Karas \& Lanza}{1998}]{akl}
   Abramowicz M. A., Karas, V., Lanza.,A. 1998, A\&A, 331, 1143

\bibitem[\protect\citeauthoryear{Abramowicz \& Klu\'zniak}{2001}]{ak01}
   Abramowicz M. A., Klu\'zniak W. 2001 A\&A, 374, L19

\bibitem[\protect\citeauthoryear{Abramowicz, Bulik, Bursa \&
   Klu\'zniak}{2003}]{abbk} Abramowicz M. A., Bulik T., Bursa M.,
   Klu\'zniak W. 2003 A\&A, 404, L21

\bibitem[\protect\citeauthoryear{Abramowicz et
   al.}{2003}]{akklr03}Abramowicz, M. A., Karas, V., Klu\'{z}niak, W.,
   Lee, W.H., Rebusco, P. 2003, PASJ, 55, 467

\bibitem[\protect\citeauthoryear{Balsara}{1995}]{ba}
   Balsara D. 1995, J. Comp. Phys., 121, 357

\bibitem[\protect\citeauthoryear{Blaes}{1987}]{blaes87}Blaes, O. 1987,
MNRAS, 227, 975

\bibitem[\protect\citeauthoryear{Daigne \& Mochkovitch}{1997}]{dm}
   Daigne F., Mochkovitch R. 1997, MNRAS, 285, L15

\bibitem[\protect\citeauthoryear{Daigne \& Font}{2004}]{df}
   Daigne F., Font J.A. 2004, MNRAS, 349, 841

\bibitem[\protect\citeauthoryear{Davies, Benz, Piran \&
   Thielemann}{1994}]{da94} Davies M.B., Benz W., Piran T., Thielemann
   F.K. 1994, ApJ, 431, 742

\bibitem[\protect\citeauthoryear{Font \& Daigne}{2002}]{fd1}
   Font J.A., Daigne F. 2002, MNRAS, 334, 383

\bibitem[\protect\citeauthoryear{Fragile}{2005}]{fragile05}Fragile,
P. C. 2005, to appear in the proceedings of the XXII Texas Symposium
on Relativistic Astrophysics and Cosmology (astro-ph/0503305)

\bibitem[\protect\citeauthoryear{Hjorth et al.}{2003}]{hjorth03}
Hjorth, J. et al. 2003, Nature, 423, 847

\bibitem[\protect\citeauthoryear{Igumenshchev, Chen \&
   Abramowicz}{1996}]{ica} Igumenshchev I., Chen X., Abramowicz,
   M. 1996, MNRAS, 278, 236

\bibitem[\protect\citeauthoryear{Kato}{1998}]{kato98} Kato S., Fukue
   J., Mineshige S. 1998 \emph{Black Hole Accretion Disks} Kyoto
   University Press, Japan.

\bibitem[\protect\citeauthoryear{Kato}{2001}]{kato01}
   Kato S. 2001, PASJ, 53, 1 

\bibitem[\protect\citeauthoryear{Klu\'zniak \& Lee}{1998}]{kl98}
   Klu\'zniak W., Lee W. H. 1998, ApJ, 494, L53

\bibitem[\protect\citeauthoryear{Klu\'zniak et al.}{2004}]{k04}
   Klu\'zniak W., Abramowicz M. A., Kato S., Lee W. H., Stergioulas N.
   2004, ApJ, 603, L89

\bibitem[\protect\citeauthoryear{Kulkarni et
al.}{1998}]{ketal98}Kulkarni, S. et al. 1998, Nature, 395, 663

\bibitem[\protect\citeauthoryear{Lee}{2001}]{lee}
   Lee W. H. 2001, MNRAS, 328, 583

\bibitem[\protect\citeauthoryear{Lee \& Ramirez--Ruiz}{2002}]{leerm}
   Lee W. H., Ramirez--Ruiz E. 2002, ApJ, 577, 893

\bibitem[\protect\citeauthoryear{Lee, Abramowicz \&
   Klu\'zniak}{2004}]{lak} Lee W. H., Abramowicz M., Klu\'zniak W.
   2004, ApJ, 603, 93L

\bibitem[\protect\citeauthoryear{McClintock \&
   Remillard}{2004}]{mr04} McClintock, J. E., Remillard, R. 2004, in
   Compact Stellar X-Ray Sources, ed. W. H. G. Lewin \& M. van der Klis
   (Cambridge: Cambridge Univ. Press), in press (astro-ph/0306213)

\bibitem[\protect\citeauthoryear{Monaghan}{1992}]{mo}
   Monaghan J. J.  1992, ARA\&A, 30, 543

\bibitem[\protect\citeauthoryear{Monaghan \& Lattanzio}{1985}]{ml}
   Monaghan J. J., Lattanzio J. C.  1985, A\&A, 149, 135

\bibitem[\protect\citeauthoryear{Masuda, Nishida \& Eriguchi}{1998}]{mne}
   Masuda N., Nishida S., Eriguchi Y. 1998, MNRAS, 297, 1139

\bibitem[\protect\citeauthoryear{Michel}{1972}]{mi}
   Michel F. 1972, AP\&SS, 15, 153

\bibitem[\protect\citeauthoryear{Montero, Rezzolla \&
Yoshida}{2004}]{mry}Montero, P., Rezzolla, L., Yoshida, S. 2004,
MNRAS, 354, 1040

\bibitem[\protect\citeauthoryear{Nomoto, Maeda, Mazzali, Umeda, Deng
   \& Iwamoto}{2004}]{no04} Nomoto K., Maeda K., Mazzali P., Umeda H.,
   Deng J., Iwamoto K. 2004, in \emph{Stellar Collapse} edited by
   Chris L. Fryer, Astrophysics And Space Science Library Vol. 302,
   Kluwer Academic Publishers, The Netherlands.

\bibitem[\protect\citeauthoryear{Ortega--Rodriguez et
al.}{2002}]{osw02}Ortega--Rodriguez, M., Silbergleit, A. S., Wagoner,
R. V. 2002, ApJ, 567, 1043

\bibitem[\protect\citeauthoryear{Paczy\'nski}{1998}]{p98}
   Paczy\'nski B. 1998, ApJ, 494, L45

\bibitem[\protect\citeauthoryear{Paczy\'nski \& Wiita}{1980}]{pw}
   Paczy\'nski B., Wiita J. 1980, A\&A, 88, 23

\bibitem[\protect\citeauthoryear{Papaloizou \& Pringle}{1984}]{pp84}
   Papaloizou J. C., Pringle J. E. 1984, MNRAS, 208, 31


\bibitem[\protect\citeauthoryear{Perez et al.}{1997}]{petal97}Perez,
C. A., Silbergleit, A. S., Wagoner, R. V., Lehr, D. E. 1997, ApJ,
476, 589

\bibitem[\protect\citeauthoryear{Rebusco}{2004}]{re}
   Rebusco P. 2004, PASJ, 56, 553

\bibitem[\protect\citeauthoryear{Rezzolla, Yoshida, Maccarone \&
   Zanotti}{2003a}]{rymz} Rezzolla L., Yoshida S., Maccarone T. J.,
   Zanotti O. 2003, MNRAS, 344, L37

\bibitem[\protect\citeauthoryear{Rezzolla, Yoshida \& Zanotti}{2003b}]{ryz}
   Rezzolla L., Yoshida S., Zanotti O. 2003, MNRAS, 344, 978

\bibitem[\protect\citeauthoryear{Rosswog, Speith \&
Wynn}{2004}]{rsw04}Rosswog, S., Speith, R., Wynn, G. A. 2004, MNRAS,
351, 1121

\bibitem[\protect\citeauthoryear{Ruffert \& Janka}{1999}]{rj99}
   Ruffert M., Janka H. Th. 1999, A\&A 344, 573

\bibitem[\protect\citeauthoryear{Ruffert, Janka \& Sch\"{a}efer}{1996}]{rjs96}
   Ruffert M., Janka H. Th., Sch\"{a}efer G. 1996, A\&A 311, 352

\bibitem[\protect\citeauthoryear{Rubio--Herrera \& Lee}{2005}]{rl}
   Rubio--Herrera E., Lee W. H. 2005, MNRAS, 357, L31

\bibitem[\protect\citeauthoryear{Schnittman}{2005}]{schnittman05}Schnittman,
J. D. 2005, ApJ, 621, 940

\bibitem[\protect\citeauthoryear{Shapiro \& Shibata}{2002}]{sasi02}
   Shapiro S., Shibata M. 2002, ApJ, 577, 904

\bibitem[\protect\citeauthoryear{Shibata \& Shapiro}{2002}]{sisa02}
   Shibata M., Shapiro S. 2002, ApJ, 572, L39

\bibitem[\protect\citeauthoryear{Silbergleit et
al.}{1999}]{swo01}Silbergleit, A. S., Wagoner, R. V.,
Ortega--Rodriguez, M. 2001, ApJ, 548, 335

\bibitem[\protect\citeauthoryear{Stanek et al.}{2003}]{setal03}Stanek,
K. et al. 2003, ApJ, 591, L17

\bibitem[\protect\citeauthoryear{van der Klis}{2000}]{vk}
   van der Klis, M. 2000, ARA\&A, 38, 717


\bibitem[\protect\citeauthoryear{van Putten}{2001}]{vp01}
   van Putten M. H. P. M. 2001, Phys. Rev. Lett. 87, No. 9. (091101--1) 

\bibitem[\protect\citeauthoryear{Wagoner}{1999}]{wagoner99}Wagoner,
R. V. 1999, Phys. Rep. 311, 259

\bibitem[\protect\citeauthoryear{Wilson}{1984}]{wi84}
   Wilson D.B. 1984, Nature, 312, 620

\bibitem[\protect\citeauthoryear{Woosley}{1993}]{w93}
   Woosley S. E. 1993, ApJ. 405, 273   

\bibitem[\protect\citeauthoryear{Zanotti, Rezzolla \&
   Font}{2003}]{zrf} Zanotti O., Rezzolla L., Font J. A. 2003, MNRAS,
   341, 832

\bibitem[\protect\citeauthoryear{Zanotti, Font, Rezzolla \&
   Montero}{2005}]{zfrm} Zanotti O., Font J. A., Rezzolla L., Montero
   P. J. 2005, MNRAS, 356, 1371

\bibitem[\protect\citeauthoryear{Zurek \& Benz}{1986}]{zb86} Zurek
   W. H., Benz W. 1986, ApJ, 308, 123

\end{thebibliography}
\end{document}